\documentclass[twocolumn, resetfootnote]{aastex701}

\def\OI{[\mbox{O\,{\sc i}}]$\lambda 6300$}
\def\OIII{[\mbox{O\,{\sc iii}}]$\lambda 5007$}

\def\OIIIab{[\mbox{O\,{\sc iii}}]$\lambda\lambda 4959,5007$}
\def\SIIab{[\mbox{S\,{\sc ii}}]$\lambda\lambda 6717,6731$}
\def\SII{[\mbox{S\,{\sc ii}}]$\lambda \lambda 6717,6731$}
\def\NII{[\mbox{N\,{\sc ii}}]$\lambda 6583$}

\def\NIIab{[\mbox{N\,{\sc  ii}}]$\lambda \lambda 6547,6583$}

\def\OI{[\mbox{O{\sc i}}]$\lambda 6300$}

\def\Ha{{H$\alpha$}}
\def\Hb{{H$\beta$}}

\def\NIIHa{[\mbox{N\,{\sc ii}}]$\lambda 6583$/H$\alpha$}
\def\SIIHa{[\mbox{S\,{\sc ii}}]$\lambda\lambda 6717,6731$/H$\alpha$}

\def\OIHa{[\mbox{O\,{\sc i}}]$\lambda 6300$/H$\alpha$}
\def\OIIIHb{[\mbox{O\,{\sc iii}}]$\lambda 5007$/H$\beta$}

\def\LOIIIs4{$L[\mbox{O\,{\sc iii}}]$/$\sigma^4$}

\def\ergs{${\rm erg}~{\rm s}^{-1}$}
\def\kms{${\rm km}~{\rm s}^{-1}$}

\newcommand{\ergcms}	{\ifmmode {\rm erg\,cm}^{-2}\,{\rm s}^{-1} \else erg\,cm$^{-2}$\,s$^{-1}$\fi}

\definecolor{myblue}{RGB}{0, 100, 220}
\definecolor{myred}{RGB}{225, 0, 100}

\usepackage{threeparttable} % in the preamble
\usepackage{gensymb}
\usepackage{graphicx}
\usepackage{sistyle}
\SIthousandsep{,}
\usepackage[utf8]{inputenc}
\DeclareUnicodeCharacter{202F}{\,}

\received{April 20, 2026}
\revised{June 11, 2026}
\accepted{June 12, 2026}
\journalinfo{}
\begin{document}

\title{Hector Galaxy Survey: Optical IFU and Chandra Reveal a Low-Luminosity AGN Behind Extended LINER Emission}
\correspondingauthor{Kyuseok Oh}
\email[show]{oh@kasi.re.kr}

\author[orcid=0000-0002-5037-951X]{Kyuseok Oh}
\affiliation{Korea Astronomy and Space Science Institute, Daedeok-daero 776, Yuseong-gu, Daejeon 34055, Republic of Korea}
\email[]{oh@kasi.re.kr}  

\author[orcid=0009-0009-9074-716X]{Gabriella Quattropani}
\affiliation{School of Mathematical and Physical Sciences, Macquarie University, Sydney, NSW 2109, Australia}
\affiliation{Astrophysics and Space Technologies Research Centre, Macquarie University, Sydney, NSW 2109, Australia}
\affiliation{ARC Centre of Excellence for All Sky Astrophysics in 3 Dimensions (ASTRO 3D), Australia}
\email[]{gabriella.quattropani@hdr.mq.edu.au}

\author[orcid=0000-0002-4326-8598]{Pratyush Kumar Das}
\affiliation{School of Mathematics and Physics, University of Queensland, Brisbane, QLD 4072, Australia}
\email[]{p.das@uq.edu.au}

\author[orcid=0000-0002-3032-2292]{Minje Beom}
\affiliation{Korea Astronomy and Space Science Institute, Daedeok-daero 776, Yuseong-gu, Daejeon 34055, Republic of Korea}
\email[]{minje.beom@gmail.com}

\author[orcid=0000-0003-3451-0925]{Joon Hyeop Lee}
\affiliation{Korea Astronomy and Space Science Institute, Daedeok-daero 776, Yuseong-gu, Daejeon 34055, Republic of Korea}
\email[]{frozfire@gmail.com}

\author[orcid=0000-0002-1045-2559]{Oğuzhan Çakır} %O\u{g}uzhan \c{C}ak{\i}r
\affiliation{School of Mathematical and Physical Sciences, Macquarie University, Sydney, NSW 2109, Australia}
\affiliation{Astrophysics and Space Technologies Research Centre, Macquarie University, Sydney, NSW 2109, Australia}
\affiliation{ARC Centre of Excellence for All Sky Astrophysics in 3 Dimensions (ASTRO 3D), Australia}
\email[]{oguzhan.cakir@students.mq.edu.au}

\author[orcid=0000-0003-2723-0810]{Andrei Ristea}
\affiliation{Centre for Astrophysics and Supercomputing, Swinburne University of Technology, Hawthorn, VIC 3122, Australia}
\affiliation{ARC Centre of Excellence in Optical Microcombs for Breakthrough Science (COMBS)}
\email[]{aristea@swin.edu.au}

\author[orcid=0000-0002-0145-9556]{Hyunjin Jeong}
\affiliation{Korea Astronomy and Space Science Institute, Daedeok-daero 776, Yuseong-gu, Daejeon 34055, Republic of Korea}
\email[]{hyunjin@kasi.re.kr}

\author[orcid=0000-0003-0469-345X]{Jiwon Chung}
\affiliation{Korea Astronomy and Space Science Institute, Daedeok-daero 776, Yuseong-gu, Daejeon 34055, Republic of Korea}
\email[]{jiwon@kasi.re.kr}

\author[orcid=0000-0002-7998-9581]{Michael J. Koss}
\affiliation{Eureka Scientific, Inc., 2452 Delmer Street, Suite 100, Oakland, CA 94602-3017, USA}
\email[]{Mike.Koss@eurekasci.com}

\author[orcid=0000-0003-2880-9197]{Scott M. Croom}
\affiliation{Sydney Institute for Astronomy, School of Physics, A28, The University of Sydney, Sydney, NSW 2006, Australia}
\affiliation{ARC Centre of Excellence for All Sky Astrophysics in 3 Dimensions (ASTRO 3D), Australia}
\email[]{scott.croom@sydney.edu.au}

\author{Mina Pak}
\affiliation{Korea Astronomy and Space Science Institute, Daedeok-daero 776, Yuseong-gu, Daejeon 34055, Republic of Korea}
\email[]{minapak@kasi.re.kr}

\author[orcid=0000-0002-2879-1663]{Matt S. Owers}
\affiliation{School of Mathematical and Physical Sciences, Macquarie University, Sydney, NSW 2109, Australia}
\affiliation{Astrophysics and Space Technologies Research Centre, Macquarie University, Sydney, NSW 2109, Australia}
\affiliation{ARC Centre of Excellence for All Sky Astrophysics in 3 Dimensions (ASTRO 3D), Australia}
\email[]{matt.owers@mq.edu.au}

\author[orcid=0000-0002-1576-2505]{Sarah M. Sweet}
\affiliation{School of Mathematics and Physics, University of Queensland, Brisbane, QLD 4072, Australia}
\affiliation{ARC Centre of Excellence for All Sky Astrophysics in 3 Dimensions (ASTRO 3D), Australia}
\email[]{s.sweet@uq.edu.au}

\author[orcid=0000-0003-0283-8352]{Jong Chul Lee}
\affiliation{Korea Astronomy and Space Science Institute, Daedeok-daero 776, Yuseong-gu, Daejeon 34055, Republic of Korea}
\email[]{jclee@kasi.re.kr}

\author[orcid=0000-0003-1627-9301]{Julia J. Bryant}
\affiliation{Sydney Institute for Astronomy, School of Physics, A28, The University of Sydney, Sydney, NSW 2006, Australia}
\affiliation{Astralis-USydney, Sydney Institute for Astronomy (SIfA), School of Physics, The University of Sydney, NSW, 2006, Australia}
\affiliation{ARC Centre of Excellence for All Sky Astrophysics in 3 Dimensions (ASTRO 3D), Australia}
\email[]{julia.bryant@sydney.edu.au}

\author[orcid=0000-0002-4731-9604]{Sree Oh}
\affiliation{Department of Astronomy and Yonsei University Observatory, Yonsei University, 50 Yonsei-ro, Seodaemun-gu, Seoul 03722, Republic of Korea}
\email[]{sreemario@gmail.com}

\author[orcid=0000-0003-2552-0021]{Jesse van de Sande}
\affiliation{School of Physics, University of New South Wales, Sydney, NSW 2052, Australia}
\email[]{j.van_de_sande@unsw.edu.au}

\author[orcid=0000-0002-1333-147X]{Peixin Zhu}
\affiliation{Center for Astrophysics $\vert$ Harvard \& Smithsonian, 60 Garden Street, Cambridge, MA 02138, USA}
\email[]{peixin.zhu@cfa.harvard.edu}

\author[orcid=0000-0002-9332-5386]{Stefania Barsanti}
\affiliation{Sydney Institute for Astronomy, School of Physics, A28, The University of Sydney, Sydney, NSW 2006, Australia}
\email[]{stefania.barsanti@sydney.edu.au}

\author[orcid=0000-0002-7301-461X]{Madusha L. P. Gunawardhana}
\affiliation{Sydney Institute for Astronomy, School of Physics, A28, The University of Sydney, Sydney, NSW 2006, Australia}
\affiliation{ARC Centre of Excellence for All Sky Astrophysics in 3 Dimensions (ASTRO 3D), Australia}
\email[]{madusha.gunawardhana@sydney.edu.au}

\author[orcid=0009-0002-8534-5077]{Sujeeporn Tuntipong}
\affiliation{Sydney Institute for Astronomy, School of Physics, A28, The University of Sydney, Sydney, NSW 2006, Australia}
\affiliation{ARC Centre of Excellence for All Sky Astrophysics in 3 Dimensions (ASTRO 3D), Australia}                
\email[]{stun4076@uni.sydney.edu.au}

\author{Robert Content}
\affiliation{Astralis-AAO, Australian Astronomical Optics, Faculty of Science and Engineering, Macquarie University, Sydney, NSW, Australia}
\email[]{contentrobert0@gmail.com}

\author[orcid=0000-0002-6998-6993]{Jon Lawrence}
\affiliation{Astralis-AAO, Australian Astronomical Optics, Faculty of Science and Engineering, Macquarie University, Sydney, NSW, Australia}
\email[]{jon.lawrence@mq.edu.au}

\author{Ayoan Salim Sadman}
\affiliation{School of Aerospace, Mechanical and Mechatronic Engineering, Faculty of Engineering, The University of Sydney, NSW, Australia}
\affiliation{Astralis-USydney, Sydney Institute for Astronomy (SIfA), School of Physics, The University of Sydney, NSW, 2006, Australia}
\email[]{ayoansalim@gmail.com}

\author{Will Saunders}
\affiliation{Astralis-AAO, Australian Astronomical Optics, Faculty of Science and Engineering, Macquarie University, Sydney, NSW, Australia}
\email[]{will.saunders@mq.edu.au}

\author[orcid=0000-0002-8352-7515]{Barnaby Norris}
\affiliation{Sydney Institute for Astronomy, School of Physics, A28, The University of Sydney, Sydney, NSW 2006, Australia}
\affiliation{Astralis-USydney, Sydney Institute for Astronomy (SIfA), School of Physics, The University of Sydney, NSW, 2006, Australia}
\email[]{barnaby.norris@sydney.edu.au}

\author{Gurashish Singh Bhatia}
\affiliation{ICRAR, University of Western Australia, 35 Stirling Highway, Crawley, WA 6009, Australia}
\affiliation{Astralis-USydney, Sydney Institute for Astronomy (SIfA), School of Physics, The University of Sydney, NSW, 2006, Australia}
\email[]{g.ashish.singh9@gmail.com}

\begin{abstract}
We present evidence that the Hector Galaxy Survey galaxy C901005481609968 ($z_{\rm cl}=0.0553$), 
which exhibits spatially extended LINER-like emission in optical integral-field spectroscopy (IFS), 
hosts a low-luminosity active galactic nucleus (LLAGN) that contributes substantially to its ionization budget. 
Although the galaxy is not selected as an AGN by mid-infrared AGN color criteria, 
archival Chandra data reveal a compact nuclear X-ray source with 
$\log L_{\rm X}\approx41.46$\,\ergs, supporting the presence of an LLAGN. 
Spatially resolved emission-line diagnostics show LINER-like line ratios across most spaxels with $\mathrm{S/N} \geq 3$, 
while spatially resolved $\tau$ maps ($\tau \equiv Q_{\rm pAGB}/Q_{\rm req}$) indicate a widespread photon deficit 
($\log\tau<0$ over most of the mapped region), even under the most optimistic pAGB normalizations, 
the nuclear region remains at $\tau < 1$.  
Line-ratio--kinematic tests find no evidence for shock-dominated excitation as the primary driver of the extended emission, although a localized or sub-dominant shock contribution cannot be ruled out with the present data. 
We use this galaxy as a pilot case because the combination of Hector IFS and an independent nuclear X-ray constraint 
provides a stringent validation of the spatially resolved photon-budget framework. 
Our results indicate that evolved stellar populations alone cannot account for the observed emission, 
that an additional nuclear ionizing source is required at least in the inner region, 
and that a weak LLAGN likely contributes to the ionizing budget, particularly in the inner region. 
Our results demonstrate that extended LINER-like emission can conceal a substantial LLAGN contribution 
even when traditional optical and infrared AGN indicators are weak, 
and that spatially resolved photon-budget tests combined with X-ray constraints can effectively reveal such hidden activity.
\end{abstract}



\section{Introduction} 

Low-ionization nuclear emission-line regions (LINERs) were first recognized as a distinct class of 
galactic nuclei in the early 1980s \citep{Heckman80, Baldwin81}. 
They are characterized by prominent low-ionization emission lines (e.g., \OI, \NII, and \SIIab) 
relative to the Balmer emission lines, and have since attracted sustained attention as probes of nuclear activity.
LINER-like emission is now known to be common in the nuclei of massive, 
early-type galaxies in the local Universe \citep{Ho97b}, 
and is frequently identified using standard optical emission-line Baldwin-Phillips-Terlevich (BPT) diagnostic diagrams 
\citep[e.g.,][]{Baldwin81, Veilleux87, Kauffmann03, Kewley01, Kewley06, Schawinski07}.

Over the past four decades, considerable effort has been devoted to identifying the primary ionization mechanism responsible for LINER emission. 
Proposed scenarios include shock heating in galactic winds or accretion flows \citep{Heckman80, Dopita95}, 
photoionization by young, hot stars in star-forming regions \citep{Terlevich85}, 
and low-luminosity active galactic nuclei (LLAGN) producing a hard ionizing spectrum \citep{Ferland83, Halpern83}. 
The ambiguity arises from the fact that LINER-like line ratios can result from multiple excitation processes under different physical conditions.

The advent of wide-area spectroscopic surveys, such as the Sloan Digital Sky Survey \citep[SDSS;][]{York00}, 
together with modern integral-field spectroscopic surveys including 
Spectrographic Areal Unit for Research on Optical Nebulae \citep[SAURON;][]{Bacon01}, 
Calar Alto Legacy Integral Field Area Survey \citep[CALIFA;][]{Sanchez12}, 
Sydney-AAO Multi-object Integral Field Spectrograph Survey \citep[SAMI;][]{Croom12, Bryant15}, 
and Mapping Nearby Galaxies at Apache Point Observatory \citep[MaNGA;][]{Bundy15, Drory15}, 
have revealed that LINER-like emission is frequently spatially extended over kiloparsec scales, 
in contrast to the purely nuclear confinement expected from photoionization by a compact AGN. 
This has motivated the idea that extended LINER emission may instead be powered by evolved stellar populations, 
in particular hot, low-mass evolved stars (HOLMES), including post-asymptotic giant branch (post-AGB; pAGB) stars 
\citep{Binette94, Stasinska08, Sarzi10, Singh13, Belfiore16}. 
Such stars can supply a diffuse Lyman-continuum radiation field capable of sustaining low-level nebular emission, 
especially in massive, quiescent galaxies.

Nonetheless, growing evidence suggests that some LINERs do indeed harbor accreting supermassive black holes. 
High-resolution X-ray observations have identified compact nuclear sources in a large fraction of LINER-classified galaxies, 
with reported detection rates ranging from $\sim 50\%$ in serendipitous surveys \citep{Constantin09} 
to $\sim 74$--$86\%$ in targeted Palomar samples \citep{Ho01, Ho08, GonzalezMartin09}, 
and multi-wavelength studies have shown that weak AGNs can coexist with extended low-ionization emission (e.g., \citealt{Dudik05, Eracleous10}; 
see also compact-core studies such as \citealt{Filho04}). 
In these cases, the observed line ratios alone may be insufficient to distinguish AGN excitation from stellar photoionization, 
particularly in low-luminosity systems where traditional AGN indicators such as broad-line regions, 
strong mid-infrared emission, or high-ionization UV lines are absent or faint.

Disentangling these competing scenarios benefits from instruments that combine kpc-scale spatial coverage 
with the spectral resolution required to identify the relevant low-ionization diagnostics. 
The Hector Galaxy Survey, with its dedicated multi-object IFU design and hexabundle fiber sampling, 
is well suited to providing such spatially resolved diagnostics across large samples of nearby galaxies. 
When combined with an independent high-resolution X-ray probe of the nucleus, 
it offers a natural framework for testing whether traditional optical and infrared diagnostics may miss 
substantial AGN contributions in LINER-like systems.

This ambiguity motivates a spatially resolved test of the ionizing photon budget, 
ideally paired with an independent probe of nuclear accretion. 
In this paper, we analyze a Hector Galaxy Survey galaxy that exhibits spatially extended LINER-like emission in the optical, 
yet lacks strong mid-infrared AGN signatures, 
while deep archival Chandra data reveal a compact nuclear X-ray source. 
Using Hector IFS, we apply the $\tau$ diagnostic, 
building on the ionizing photon-budget framework developed in earlier studies of LINER-like and retired galaxies 
\citep{Binette94, CidFernandes11, Papaderos13, Belfiore16, Gomes16}. 
Following \citet{Papaderos13}, we define $\tau \equiv Q_{\rm pAGB}/Q_{\rm req}$, 
where $Q_{\rm pAGB}$ is the ionizing-photon rate expected from hot evolved stars and 
$Q_{\rm req}$ is the rate required to power the observed recombination-line emission. 
In this framework, $\tau \gtrsim 1$ indicates that pAGB photoionization is sufficient in principle, 
whereas $\tau < 1$ implies a local photon deficit and therefore the need for an additional ionizing source. 
Our goal is not simply to identify LINER-like line ratios, but to test, as a function of radius, 
whether evolved stars can plausibly account for the observed emission once the photon budget is evaluated in a spatially resolved manner.

We present this system as a pilot case for validating the spatially resolved photon-budget framework 
against an independent nuclear X-ray constraint. 
The combination is particularly informative because the optical data trace the extended line-emitting gas, 
whereas the X-ray data isolate the compact accretion-powered nucleus. 
As we show below, the spatially resolved $\tau$ analysis disfavors a purely pAGB-powered interpretation 
over most of the mapped region, especially in the nucleus, 
while the Chandra detection demonstrates that an LLAGN is present and energetically relevant. 
Rather than assuming that a single mechanism dominates at all radii, 
we use this pilot case to test whether extended LINER-like emission can conceal a substantial LLAGN contribution 
even when traditional optical or infrared AGN indicators are weak.

The remainder of this paper is organized as follows. 
Section~\ref{sec:data} describes the observations, data reduction, and spectral analysis of the optical IFS data, 
as well as the processing of the archival Chandra X-ray observations. 
This section also presents the spatially resolved emission-line diagnostics, tests for shock excitation, 
and the ionizing-photon budget analysis including the $\tau$ diagnostic. 
Section~\ref{sec:discussion} discusses the implications of these results, 
including comparisons with other LINER-like systems and the evidence for an LLAGN. 
Finally, Section~\ref{sec:conclusion} summarizes our main conclusions.

Throughout this paper, we adopt a cosmology with $h = 0.70$, $\Omega_{M} = 0.30$, and $\Omega_{\Lambda} = 0.70$.

\section{Observational Analysis} \label{sec:data}

\subsection{Optical Integral-Field Spectroscopy} \label{subsec:optical}

The IFS data cube of the target galaxy was obtained 
on 1 September 2024 as part of the Hector Galaxy Survey \citep{Bryant24, OhSree25}, 
a large ongoing spectroscopic survey conducted with the 3.9-meter Anglo-Australian Telescope (AAT). 
The survey aims to obtain spatially resolved spectra for up to 15,000 galaxies at redshift $z<0.1$, 
providing a statistically representative sample of the nearby Universe. 
Observations are performed using the Hector instrument, 
which employs multiple hexabundle fiber \citep{BlandHawthorn11, Bryant14, Brown18, Wang23} integral field units (IFUs), 
each consisting of closely packed optical fibers that simultaneously collect spectra from different regions across the projected extent of a galaxy.

The observations presented in this work were obtained with the AAOmega spectrograph \citep{Sharp06}, 
which delivers spectral resolutions of $R_{\rm blue}=1882$ (Full Width at Half Maximum, FWHM = 2.55~\AA) and 
$R_{\rm red}=4507$ (FWHM = 1.52~\AA), 
with wavelength coverage of approximately 3750--5750~\AA\ in the blue arm and 6300--7400~\AA\ in the red arm.
For comparison, Hector can also observe with the higher-resolution Spector spectrograph \citep{Content18, Mohanan22, Zhelem22}, 
which provides $R_{\rm blue} = 3429$ (FWHM = 1.40\,\AA, 3750--5850~\AA) and $R_{\rm red} = 5667$ (FWHM = 1.20\,\AA, 5750--7800~\AA).

The target galaxy (Hector ID: C901005481609968; R.A.\,=\,20:13:07.83, Decl.\,=\,$-$56:48:29.78; J2000) has
an optical spectroscopic redshift of $z = 0.0597$. 
The galaxy is a confirmed member of the Abell~3667
cluster based on the membership analysis of \citet{Owers09}. 
The corresponding rest-frame velocity offset from the cluster mean redshift,
$\Delta v = c\,(z_{\rm spec} - z_{\rm cl})/(1+z_{\rm cl}) \approx 1250$~km~s$^{-1}$, corresponds to 
$\approx 1.2\,\sigma_{\rm cl}$ for a cluster velocity dispersion of $\sigma_{\rm cl} = 1056 \pm 38$~km~s$^{-1}$ \citep{Owers09}, 
well within the range typically adopted for cluster membership ($\lesssim 2$--$3\,\sigma_{\rm cl}$).
The galaxy's identification as a ram-pressure stripping (referred to as RPS hereafter) candidate 
(\c{C}ak{\i}r et~al., submitted; see also Section~\ref{subsec:assessing}) provides independent evidence for 
its interaction with the Abell~3667 intracluster medium. We adopt the cluster redshift $z_{\rm cl} = 0.0553$ \citep{Owers09} for distance-dependent 
quantities, corresponding to a luminosity distance of $\approx 246.8$~Mpc in our adopted cosmology. 
Using $z_{\rm spec} = 0.0597$ instead would increase $D_L$ by $\approx 6\%$, shifting all luminosity-based 
quantities by $\approx 0.05$~dex---well within the dominant systematic uncertainties in our analysis (e.g., the ${\sim}\,1$~dex range in $q_{\rm pAGB}$). 
The galaxy has a stellar mass of $M_{\rm *} = 10^{10.97}\,M_\odot$ and an effective radius of $R_{\rm eff,\textit{r}} = 9.53''$ 
(corresponding to $\approx 10.4$~kpc at $z_{\rm cl} = 0.0553$) in the $r$~band.

Photometric properties used to derive $M_{\rm *}$ and $R_{\rm eff, r}$ are taken from 
the internal Hector photometric catalog shared among the collaboration (M. Beom  et al., in prep.).
The photometry was performed using a multi-gaussian expansion (MGE) model with \texttt{MGEfit} \citep{Emsellem94, Cappellari02, Cappellari12},
ellipse-based aperture photometry with \texttt{photutils} \citep{Bradley25}, 
and $k$-correction and stellar mass estimation with \texttt{kcorrect} \citep{Blanton07}, 
applied to $g$, $r$, $i$, and $z$ band images from the DESI Legacy Imaging Surveys Data Release 10 \citep{Dey19}.

% ------------------------------------------------------------------------------------------------------------------------------------------------------------------
\begin{figure*}
\centering
\includegraphics[height=0.84\textheight]{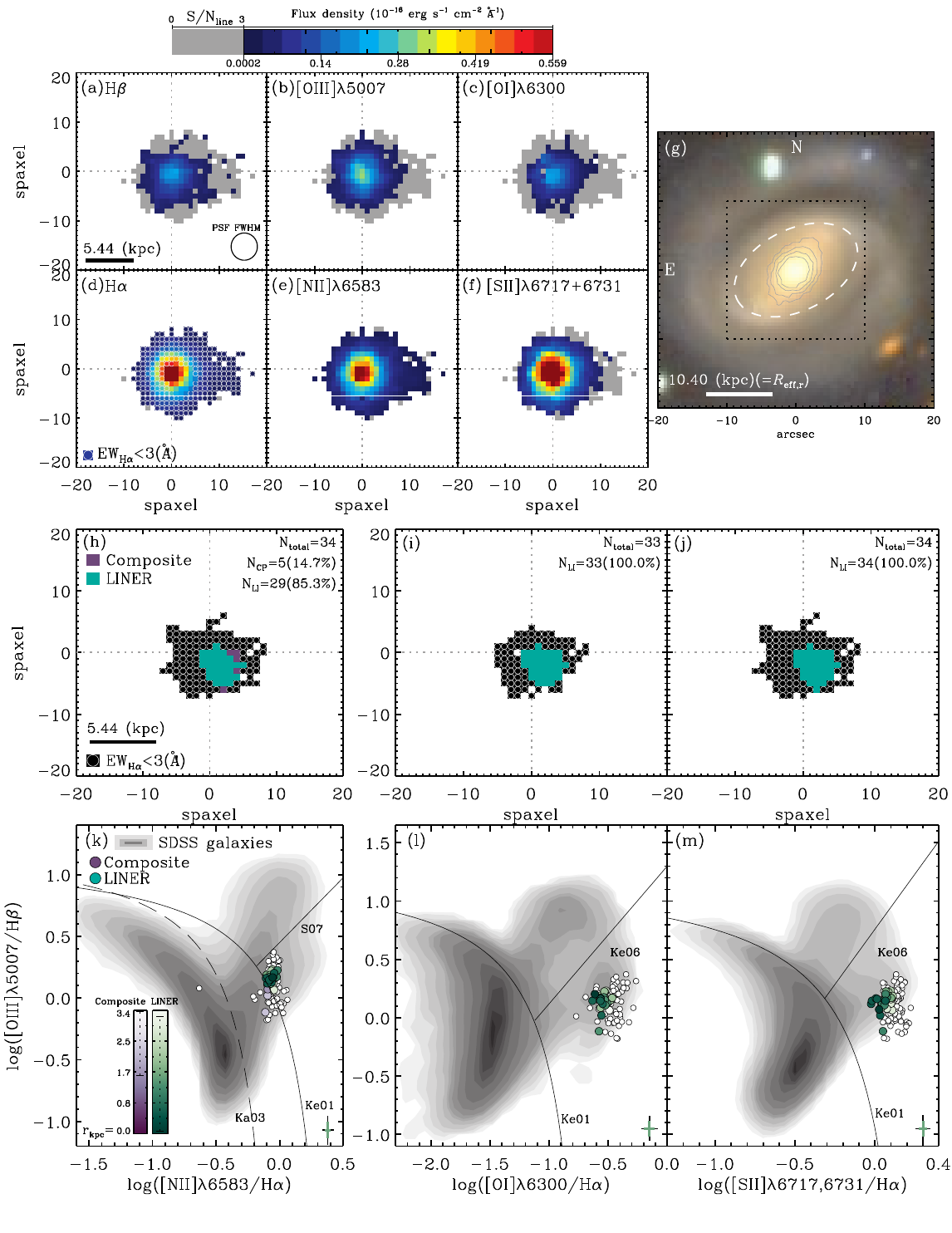}
\caption{Emission-line flux maps (a--f), a DESI Legacy Surveys DR10 \textit{grz} image (g; \citealt{Dey19}),
and BPT classification maps (h--j) and BPT diagnostic diagrams (k--m).
Panels (a--f) use a common \textit{asinh} stretch ($Q=0.001$) with limits set by the 0.1--99.9 percentile of the combined (positive) flux distribution 
across the six lines; spaxels with ${\rm S/N} \geq 3$ are colored and those with ${\rm S/N}<3$ are gray (Point Spread Function (PSF) FWHM shown in a).
Panel (g) marks the map footprint (dotted box) and $R_{\rm eff,r}$ (ellipse; P.A.=116.97\degree); contours trace the stellar continuum, highlighting the bar.
Panels (h--j) use spaxels with $\mathrm{S/N} \geq 3$ in all four BPT lines.
In (k--m), spaxels are color-coded by radius;  ${\rm EW}_{\rm H\alpha}<3$\,\AA\ spaxels are white circles, 
and a representative error bar is shown (green: ${\rm EW}_{\rm H\alpha}\ge3$\,\AA; black: ${\rm EW}_{\rm H\alpha}<3$\,\AA).
Background contours in (k--m) show SDSS galaxies from the OSSY (Oh-Sarzi-Schawinski-Yi) catalog \citep{Oh11}.}
\label{fig:hector_bpt}
\end{figure*}%  -------------------------------------------------------------------------------------------------------------------------------------------------

Emission-line fluxes were measured using the Hector spectral fitting pipeline (G. Quattropani et al., in prep.) 
following the methodology of \citet{Owers19}. 
The analysis employs the penalized pixel-fitting method (pPXF; \citealt{Cappellari04, Cappellari17, Cappellari23}) 
to model and subtract the stellar continuum. 
First, stellar and gas kinematics are derived by fitting both individual spaxels and spatially binned spectra 
using a 12th order additive polynomial as in \citet{vandeSande17}. 
The next step is to constrain a subset of simple stellar population templates 
from the high spectral resolution X-Shooter Spectral library \citep[XSL;][]{Verro22} 
to be used in the final modeling of the continuum. 
This is achieved by fitting the spatially binned spectra again with pPXF, holding the kinematics fixed, 
and using a 12th order multiplicative polynomial to correct for the effects of dust and flux calibration uncertainties \citep{Cappellari17}. 
The templates used in the fit and the corresponding weights determined by pPXF are stored. 
For the final modeling of the continuum in each individual spaxel, 
a signal-to-noise ratio (S/N) per Angstrom is measured by determining the median of the spectrum over error in a window between $4000$--$5000$~\AA\ 
divided by the square root of the pixel size. 
The S/N per Angstrom determines which templates from the pPXF fitting of the spatially binned data are used. 
If the S/N $< 5$\, \AA$^{-1}$,  an optimal template that is the weighted sum of the templates is applied. 
Otherwise, the subset of templates with non-zero weights is used. 
With templates and kinematics in hand, a final pPXF fit is performed on individual spaxels to model the continuum 
using a 12th order multiplicative polynomial. 
The continuum is subtracted from the spectrum to allow for fitting of the emission lines only. 
The emission lines are modeled with Gaussian components using a method similar to that of \texttt{LZIFU} \citep{Ho16}. 
Example spectral fits for representative spaxels are shown in Appendix~\ref{sec:bestfit} (Figure~\ref{fig:bestfit}).

% ------------------------------------------------------------------------------------------------------------------------------------------------------------------
\begin{figure*}[!t]
\centering
\includegraphics[width=1\linewidth]{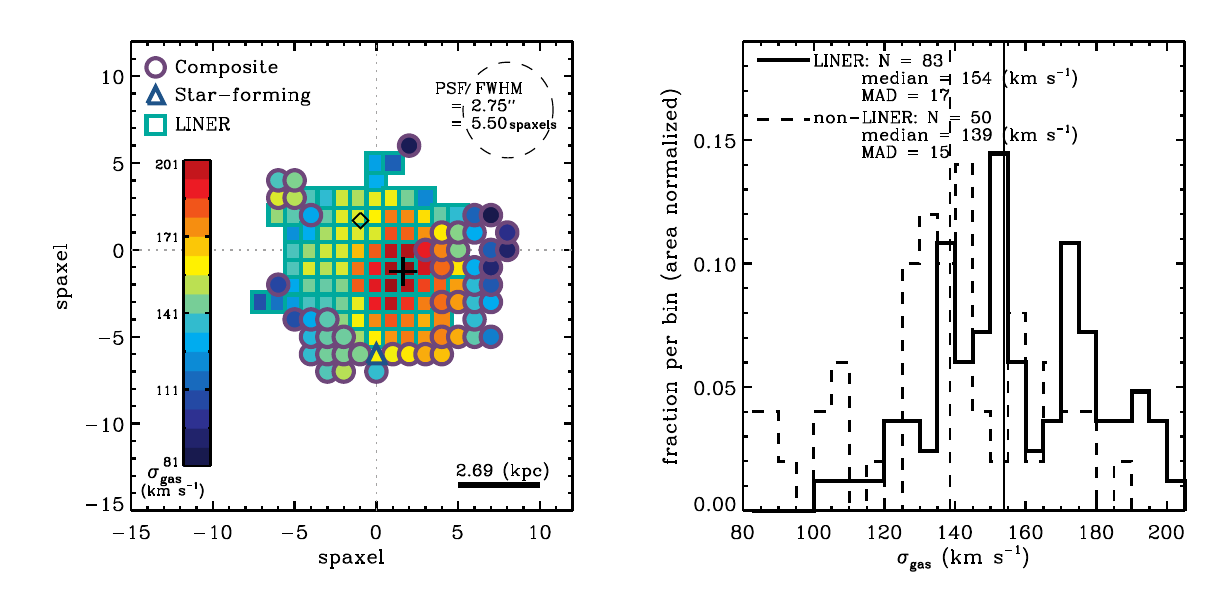}
\caption{Ionized-gas velocity dispersion and spatially resolved BPT classifications.
Left: Map of the line-of-sight ionized-gas velocity dispersion, $\sigma_{\rm gas}$ (in \kms),
from the Hector pipeline \texttt{vdisp} extension (corrected for instrumental resolution),
shown for spaxels with ${\rm S/N}_{\rm line} \geq 3$ in the BPT lines.
Symbols indicate the \NIIHa\ BPT classification: composite (purple circles), star-forming (blue triangles), and LINER-like (green squares).
The empty diamond marks the stellar light center (flux-weighted continuum centroid), and the thick cross marks the spaxel with the maximum $\sigma_{\rm gas}$.
The circle indicates the PSF FWHM ($2.75^{\prime\prime} \equiv 5.50$ spaxels).
No ${\rm EW}_{\rm H\alpha}$ cut is applied.
Right: Distributions of $\sigma_{\rm gas}$ for LINER-like (solid) and non-LINER (dashed; composite+star-forming) spaxels,
normalized to unit area, illustrating a modest shift toward higher dispersions in the LINER-like population.
Vertical lines indicate the medians of each distribution.
\label{fig:gas_velocity_dispersion}}
\end{figure*} % -------------------------------------------------------------------------------------------------------------------------------------------------

Figure~\ref{fig:hector_bpt} presents the spatially resolved emission-line properties of the target galaxy. 
Emission-line maps for \Hb, \OIII, \OI, \Ha, \NII, and \SII\ (panels a--f) show significant detections 
across an area of approximately $10^{\prime\prime} \times 10^{\prime\prime}$ ($\sim10.9\times10.9~{\rm kpc}^2$), 
with most spaxels detected at S/N$ \geq 3$. 
The emission-line morphology shows a mild asymmetry, with the line-emitting region extending preferentially 
toward the south-west, consistent with the identification of this galaxy as a RPS  
candidate in the Abell~3667 environment (\c{C}ak{\i}r et~al., submitted).
For the diagnostic analysis, we require S/N$ \geq 3$ in the relevant emission lines and an \Ha\ equivalent width 
${\rm EW}_{\rm H\alpha}\ge3$\,\AA, where ${\rm EW}_{\rm H\alpha}$ is measured with \texttt{Spaxelsleuth} (P.\ K.\ Das et al., in prep.). 
We adopt ${\rm EW}_{\rm H\alpha}\ge3$\,\AA\ to exclude very weak-emission spaxels in which line ratios are more susceptible to
ionization by hot evolved stars and measurement systematics, and to focus on regions with robust nebular emission.
Spaxels with ${\rm EW}_{\rm H\alpha}<3$\,\AA\ are shown for completeness and are explicitly marked.

For visualization, the emission-line maps in panels (a--f) are displayed using a common \textit{asinh} 
stretch ($Q=0.001$) with limits set by the $0.1$--$99.9$ percentile of the combined positive flux distribution 
across the six lines, and spaxels with ${\rm S/N}<3$ in the corresponding line are shown in gray.
In contrast, the spatial BPT-classification maps (panels h--j) and the BPT diagnostic diagrams (panels k--m) use the same diagnostic spaxel set, 
requiring S/N$ \geq 3$ in all four lines entering each diagram; spaxels are color-coded by projected distance from the galaxy center. 
Background contours indicate SDSS galaxies from the OSSY catalog at $z<0.2$ with ${\rm S/N}_{\rm line} \geq 3$ \citep{Oh11}.

The BPT diagnostics (panels h--m; \NIIHa, \OIHa, and \SIIHa\ versus \OIIIHb;
\citealt{Baldwin81, Kewley01, Kauffmann03, Kewley06, Schawinski07}) show that
the majority of spaxels fall in the LINER regime across all three diagnostics.
In the \NIIHa\ diagram, a small subset lies in the composite region,
whereas the \SIIHa\ and \OIHa\ diagrams primarily separate Seyfert and LINER classifications.
Composite-classified spaxels have lower H$\beta$ S/N than LINER-classified spaxels (median 4.3 vs.\ 7.2).
To assess whether measurement scatter near the detection threshold inflates the composite fraction at large radii,
we recomputed the composite fraction among outer spaxels ($r \ge 6$~kpc) after requiring H$\beta$ S/N$ \geq 5$ (instead of $ \geq 3$).
The fraction changes negligibly (0.65 $\rightarrow$ 0.64), indicating that low-S/N H$\beta$ scatter is not the dominant driver of the composite excess at large radii.
When all spaxels are shown without applying an ${\rm EW}_{\rm H\alpha}$ cut,
we identify a single spaxel in the star-forming region of the \NIIHa\ diagram,
but it has ${\rm EW}_{\rm H\alpha}<3$\,\AA\ and is therefore not treated as evidence for ongoing star formation.
Notably, LINER-classified spaxels are not confined to the nucleus but extend across the IFU field of view.
The projected-radius distribution is consistent across the three BPT diagnostics,
with median radii of $r_{\rm med}=1.70~{\rm kpc}$ for \NIIHa\ ($N=29$) and \OIHa\ ($N=33$),
and $r_{\rm med}=1.94~{\rm kpc}$ for \SIIHa\ ($N=34$),
with a common 16th--84th percentile range of $r \simeq 0.76$--$2.71~{\rm kpc}$.

To test whether the LINER-like excitation identified in the spatially resolved BPT analysis is associated with dynamically broadened ionized gas,
we examined the emission-line kinematics from the Hector pipeline \texttt{vdisp} extension, which provides the line-of-sight velocity dispersion of the ionized gas
(hereafter $\sigma_{\rm gas}$) corrected for instrumental resolution (Figure~\ref{fig:gas_velocity_dispersion}).
The left panel shows the spatial distribution of $\sigma_{\rm gas}$ for spaxels classified in the \NIIHa\ BPT diagnostic diagram, 
while the right panel compares the $\sigma_{\rm gas}$ distributions of LINER-like and non-LINER spaxels.
For the AAOmega setup used here, the instrumental resolution corresponds to $\sigma_{\rm inst}\approx 68~{\rm km~s^{-1}}$ in the blue arm ($R_{\rm blue}=1882$)
and $\sigma_{\rm inst}\approx 28~{\rm km~s^{-1}}$ in the red arm ($R_{\rm red}=4507$), assuming a Gaussian line-spread function.
We compare $\sigma_{\rm gas}$ between spaxels classified as LINER-like and those classified as non-LINER (i.e., star-forming or composite)
in the \NIIHa\ BPT diagnostic diagram, adopting a uniform ${\rm S/N}_{\rm line} \geq 3$ threshold and imposing no EW cut on \Ha.
LINER-like spaxels show moderately higher dispersions,
with a median $\sigma_{\rm gas}=154~{\rm km~s^{-1}}$ (Median Absolute Deviation, MAD $=17~{\rm km~s^{-1}}$),
compared to $\sigma_{\rm gas}=139~{\rm km~s^{-1}}$ (MAD $=15~{\rm km~s^{-1}}$) for non-LINER spaxels.
For context, $\sigma_\star = 174 \pm 2~{\rm km~s^{-1}}$ within $R_{\rm eff}$ (J. H. Lee et al., in prep.), 
slightly higher than the median $\sigma_{\rm gas}$ in both subsets.
This offset suggests that LINER-like regions are preferentially associated with modestly enhanced line widths,
potentially reflecting additional dynamical heating or unresolved multi-component structure.
However, caution is warranted when interpreting $\sigma_{\rm gas}$, as it can be affected by beam smearing, unresolved line profiles,
and non-circular motions.
Accordingly, we treat the dispersion comparison as qualitative, and rely primarily on the ionizing photon budget and
$\tau$ diagnostics (Sections~\ref{subsec:budget} and \ref{subsec:tau}) to constrain the excitation mechanism.

Motivated by the spatial concentration of high-$\sigma_{\rm gas}$ spaxels, 
we also tested whether the dispersion enhancement is aligned with the stellar light center.
We estimated the stellar center as the flux-weighted centroid of a wavelength-integrated continuum image (open diamond) and 
marked the spaxel with the maximum $\sigma_{\rm gas}$ (thick cross).
To quantify the location of the broader high-dispersion region (rather than a single-spaxel peak), 
we additionally computed a $\sigma_{\rm gas}$-weighted centroid of the highest-dispersion spaxels (top 5\% in $\sigma_{\rm gas}$), 
using the same spaxel selection as shown in the map.
The resulting high-$\sigma_{\rm gas}$ centroid is offset from the stellar center by $1.96^{\prime\prime}$ ($\approx 2.11$~kpc), 
indicating that the dispersion enhancement is not perfectly co-spatial with the stellar light peak.
Because the measured offset depends on the adopted mask and centroid definition, we treat this comparison as qualitative.

% ------------------------------------------------------------------------------------------------------------------------------------------------------------------
\begin{figure}
\centering
\includegraphics[width=1\linewidth, angle=0]{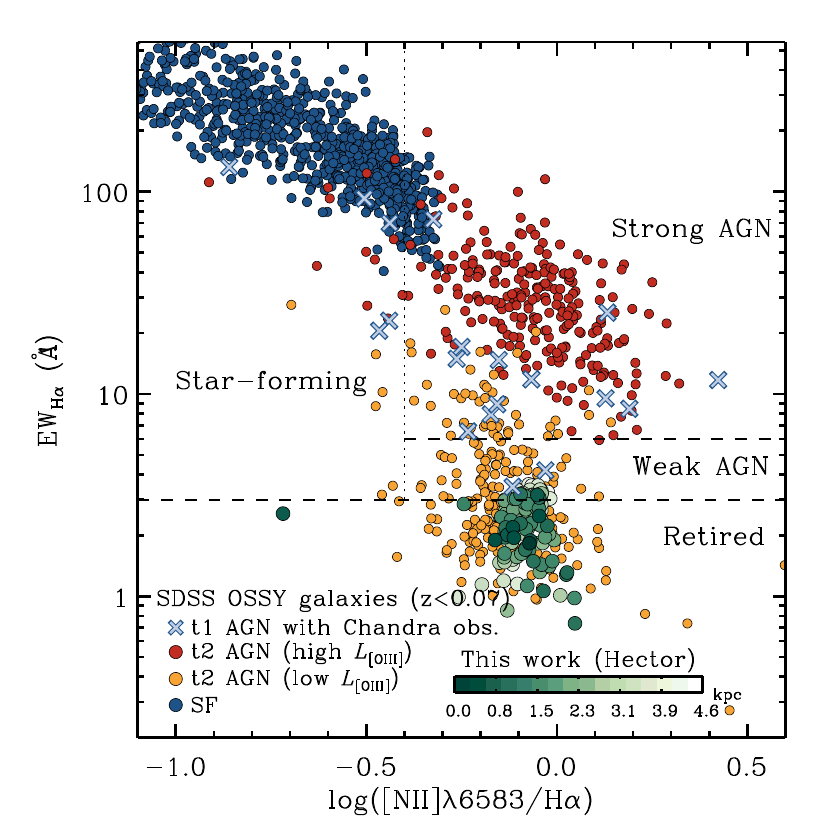}
\caption{Emission-line properties in the WHAN diagram. 
Spaxels are color-coded by projected distance from the galaxy center.
For comparison, we show SDSS galaxies and AGNs from the OSSY catalog at $z<0.07$, 
adopting the OSSY measurements and spectral classifications \citep{Oh11, Oh15}.
Crosses denote type~1 AGNs with Chandra detections (light-blue).
Filled circles show
type~2 AGNs with high \OIII\ luminosity 
(red; 10\%; $\overline{\log L_{\rm [OIII]}} \sim 41.2$~\ergs), 
type~2 AGNs with low \OIII\ luminosity 
(light-orange; 10\%; $\overline{\log L_{\rm [OIII]}} \sim 39.0$~\ergs), 
and star-forming galaxies 
(blue; 1\%; $\overline{\log {\rm SFR}} \sim 1.0~M_\odot\,{\rm yr}^{-1}$).
The displayed type~2 AGNs in each \OIII-luminosity regime (10\%) and star-forming galaxies (1\%) are 
randomly selected from their respective parent samples to reduce overplotting.
\label{fig:whan}}
\end{figure} % -------------------------------------------------------------------------------------------------------------------------------------------------

%  ------------------------------------------------------------------------------------------------------------------------------------------------------------------
\begin{figure*}
\centering
\includegraphics[width=1.0\linewidth, angle=0]{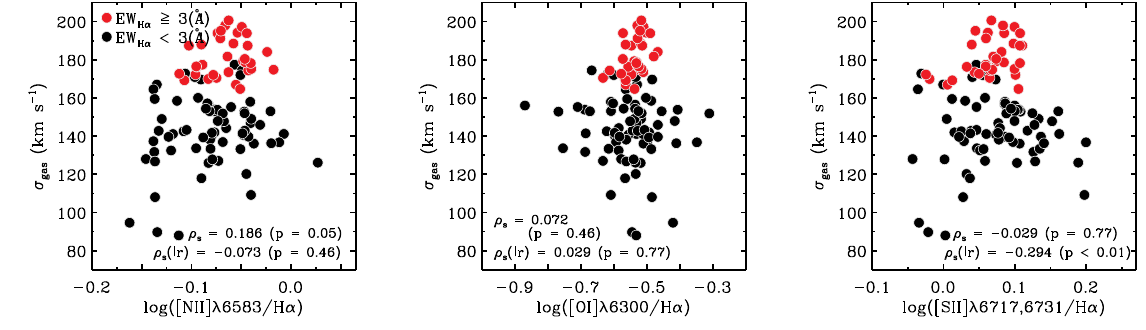}
\caption{Line ratio versus velocity dispersion test for shock excitation.
Each panel shows the spaxel-by-spaxel relation between the ionized-gas velocity dispersion
$\sigma_{\rm gas}$ and a low-ionization line ratio: $\log$\NIIHa\ (left),
$\log$\OIHa\ (middle), and $\log$\SIIHa\ (right).
Spaxels are required to have ${\rm S/N} \geq 3$ in the relevant emission lines.
Each panel reports Spearman's rank correlation coefficient $\rho_s$ (two-sided $p$-value)
and the partial Spearman coefficient $\rho_s(\,|\;r)$ (two-sided $p$-value) after controlling for projected radius $r$.
High-EW spaxels (${\rm EW}_{\rm H\alpha}\ge3$\, \AA) are shown in red, and low-EW spaxels (${\rm EW}_{\rm H\alpha}< 3$\, \AA) are shown in black.
\label{fig:shock_check}}
\end{figure*}
% -------------------------------------------------------------------------------------------------------------------------------------------------

The WHAN diagram ($W_{\rm H\alpha}$ versus \NIIHa; hereafter WHAN) combines 
\NIIHa\ and ${\rm EW}_{\rm H\alpha}$ to separate star-forming and AGN-like excitation from low-ionization emission 
associated with HOLMES, and is particularly useful when BPT classifications become uncertain due to weak \Ha\ emission \citep{CidFernandes11}. 
We use WHAN as a complementary diagnostic to assess whether the widespread LINER-like line ratios are accompanied by low ${\rm EW}_{\rm H\alpha}$ values 
expected for “retired” (HOLMES-dominated) ionization.

In the WHAN diagram (Figure~\ref{fig:whan}), 
only a small number of spaxels have ${\rm EW}_{\rm H\alpha}\ge3$\,\AA\ and lie near the boundary between weak AGNs and retired systems. 
The majority of spaxels fall below this threshold and occupy the retired region, 
consistent with a substantial contribution from evolved stellar populations, especially over much of the galaxy.

However, the WHAN classification alone cannot unambiguously determine the dominant ionizing mechanism. 
In particular, low ${\rm EW}_{\rm H\alpha}$ can reflect not only ionization by evolved stars but also 
geometric effects such as a low gas covering fraction or Lyman-continuum photon escape, 
and LINER-like line ratios are not unique to a single excitation mechanism.
A quantitative assessment of the ionizing photon budget is needed to test whether old stellar populations can energetically account for the observed emission. 
Before turning to the photon-budget analysis, we first examine whether shocks could plausibly contribute to the low-ionization excitation, 
since shocks can produce LINER-like line ratios and are often associated with broadened emission lines \citep[e.g.,][]{Allen08, Rich10}. 
We then characterize the nuclear source using the Chandra data in Section~\ref{subsec:xray}, 
and finally return to the ionizing-photon budget in Section~\ref{subsec:budget}.

We apply the ${\rm EW}_{\rm H\alpha}\ge3$\,\AA\ criterion only for visualizing high-confidence line-ratio spaxels (Figure~\ref{fig:hector_bpt}), 
but do not impose it for integrated luminosity or kinematic analyses, as the extended emission component predominantly has low EW. 
Accordingly, in the shock test below we analyze both the full spaxel set (no EW cut) and the high-EW subset, 
to assess whether any apparent line-ratio--$\sigma_{\rm gas}$ trends are driven by selection effects.

\subsection{Testing the shock-dominated scenario}
\label{subsec:shock}

To evaluate whether shocks contribute to the extended LINER-like emission, 
we examined spaxel-by-spaxel correlations between low-ionization line ratios and the ionized-gas
velocity dispersion, $\sigma_{\rm gas}$. In shock-dominated regions, elevated low-ionization ratios
(e.g., \OIHa\ and \SIIHa) are expected to be associated with broader emission lines, 
as shocks simultaneously enhance collisional excitation and inject kinetic energy
into the ionized gas \citep[e.g.,][]{Dopita95, Dopita96}.

Throughout this work, ``shock excitation'' refers to emission from gas heated and ionized in fast radiative shocks,
regardless of the physical driver (e.g., star-formation-driven winds or supernova feedback, AGN-driven outflows/jets,
bar-driven inflows, or environmental processes such as RPS).
Our goal is not to exclude shocks altogether, but to test whether radiative shocks are required to explain,
or can dominate, the extended LINER-like emission.
Accordingly, our tests are designed to assess whether radiative shocks are energetically and spatially dominant,
rather than whether shocks exist at any level.

Line-ratio--$\sigma_{\rm gas}$ trends have been used as empirical shock diagnostics in IFS studies of
galactic winds and mergers, where shocked spaxels often exhibit LINER-like ratios together with elevated
velocity dispersions \citep[e.g.,][]{Rich10, Rich11}. 
More generally, shock excitation in wind-driven systems is often inferred from a combination of spatially resolved line ratios,
gas kinematics, and geometric context, rather than from any single diagnostic alone \citep[e.g.,][]{Fogarty12}. 
Interpreting such behavior is also aided by comparisons to radiative shock models and shock and precursor model grids,
which predict systematic increases in low-ionization line ratios with shock velocity \citep[e.g.,][]{Allen08}.

As a complementary qualitative check, we overplotted the \citet{Allen08} radiative shock grids on the \NIIHa\ BPT diagnostic diagram 
(Appendix~\ref{sec:shock_models}; Figure~\ref{fig:allen08_grid}). 
While some portions of the grid overlap the LINER-like locus, 
matching the Hector spaxels would require fine-tuned combinations of shock velocity, magnetic parameter, and precursor contribution, 
and the same configuration does not simultaneously provide a consistent explanation across the explored density variants. 
Given these degeneracies, we adopt the line-ratio--$\sigma_{\rm gas}$ relations as the primary empirical discriminator of shock dominance.

We note that all line ratios and $\sigma_{\rm gas}$ measurements used in this section are based on single-component Gaussian fits. 
Multi-component decomposition (e.g., narrow + broad) can sharpen shock or outflow signatures 
in some systems \citep[e.g.,][]{Oh24, Oh25}. 
However, the Hector spectra used here do not show statistically significant non-Gaussian wings 
or systematic residual structure that would require additional components at the adopted S/N thresholds. 
We therefore restrict our analysis to single-component measurements.

Figure~\ref{fig:shock_check} summarizes these correlations, reporting both the Spearman rank correlation coefficient
($\rho_s$) and the partial correlation controlling for projected radius ($\rho_s(\mid r)$) to account for
radial dependencies in both line ratio and kinematics.
Controlling for radius is useful because both low-ionization line ratios and the measured $\sigma_{\rm gas}$ can vary systematically 
with distance from the center owing to large-scale kinematic gradients and observational effects.
In particular, beam smearing is typically strongest in the inner regions, where steep velocity gradients can inflate the apparent line widths.
The partial correlation tests whether any line-ratio--$\sigma_{\rm gas}$ association remains after removing these radius-driven trends.
A significant residual correlation would be consistent with additional local broadening (e.g., shocks), 
whereas the absence of such a residual trend suggests that the apparent correlation is dominated by radius-dependent covariance.

Using all spaxels with ${\rm S/N} \geq 3$ in the relevant lines (no EW cut; $N=105$),
we find no significant correlation between $\sigma_{\rm gas}$ and either
$\log([\mathrm{O\,I}]/\mathrm{H}\alpha)$
($\rho_s=0.072$; 95\% confidence interval (CI) $[-0.108,\,0.256]$; $\rho_s(\mid r)=0.029$; 95\% CI $[-0.158,\,0.222]$) or
$\log([\mathrm{N\,II}]/\mathrm{H}\alpha)$
($\rho_s=0.186$; 95\% CI $[-0.009,\,0.377]$; $\rho_s(\mid r)=-0.073$; 95\% CI $[-0.254,\,0.118]$).
For $\log([\mathrm{S\,II}]/\mathrm{H}\alpha)$, the raw correlation is consistent with zero
($\rho_s=-0.029$; 95\% CI $[-0.239,\,0.179]$), whereas the partial correlation controlling for radius is negative (95\% CI excludes zero) 
($\rho_s(\mid r)=-0.294$; 95\% CI $[-0.482,\,-0.083]$), opposite to the positive trend commonly expected for widespread shock-dominated excitation.

As a robustness check against low-S/N systematics, we repeat the analysis using a stricter cut
(S/N$ \geq 5$; $N=66$), which yields the same qualitative conclusion: the [S\,II]/H$\alpha$ partial correlation
remains negative, with $\rho_s(\mid r)=-0.39$ (95\% CI $[-0.58,-0.15]$).
While [O\,I]/H$\alpha$ shows a positive partial correlation at S/N$ \geq 5$
($\rho_s(\mid r)=0.35$; 95\% CI $[0.11,0.53]$; $p\simeq2.5\times10^{-3}$),
this signal is not mirrored by [N\,II]/H$\alpha$ or [S\,II]/H$\alpha$.
The [O\,I]\,$\lambda6300$ line is uniquely sensitive to the partially ionized zone where weak shocks
can produce enhanced low-ionization emission \citep{Dagostino19}, so this residual correlation may
indicate a localized, sub-dominant shock contribution in the highest-dispersion gas.
We note, however, that the [O\,I] partial correlation accounts for only $\rho_s^2\simeq 12\%$ of the
explained variance, and that the absence of a corresponding [S\,II]/H$\alpha$ correlation does not by
itself argue against shocks: the [S\,II] doublet saturates at $n_e\gtrsim 10^3$--$10^4$\,cm$^{-3}$ and
is intrinsically insensitive to compressed shock fronts (see Appendix~\ref{sec:sii_ratio}).
The combined behavior is therefore consistent with shocks not being the primary driver of the extended
LINER-like emission, while leaving room for a localized shock contribution that the present diagnostics
cannot exclude. 
Finally, the region with simultaneously high low-ionization ratios and moderate $\sigma_{\rm gas}$
is sparsely populated, and becomes largely empty under stricter S/N cuts. This suggests that low-S/N
incompleteness may contribute to the apparent paucity of points in that regime through selection effects.

When the analysis is restricted to high-EW spaxels (${\rm EW}_{\rm H\alpha}\ge3$~\AA; $N=33$),
we find a strong raw correlation between $\log([\mathrm{O\,I}]/\mathrm{H}\alpha)$ and $\sigma_{\rm gas}$
($\rho_s=0.529$; 95\% CI $[0.232,\,0.725]$).
After controlling for radius, the correlation weakens and becomes statistically inconclusive
($\rho_s(\mid r)=0.314$; 95\% CI $[-0.115,\,0.596]$),
indicating at most a marginal intrinsic association between \OIHa\ and $\sigma_{\rm gas}$ in these regions.
This suggests that part of the apparent correlation is driven by spatial covariance with radius,
consistent with large-scale kinematic gradients and beam-smearing effects that are commonly encountered in IFS datasets
\citep[e.g.,][]{Davies14, Schreiber18}, although localized shock-related enhancements in \OIHa\ cannot be fully excluded.
From the Hector \texttt{V} map we infer $|dv/dr|\simeq 21$--$33~{\rm km~s^{-1}~arcsec^{-1}}$.
For a PSF with ${\rm FWHM}=2.75^{\prime\prime}$ ($\sigma_{\rm PSF}=1.17^{\prime\prime}$), 
this implies a beam-smearing contribution of $\sigma_{\rm bs}\equiv |dv/dr|\,\sigma_{\rm PSF} \sim 25$--$38~{\rm km~s^{-1}}$, 
which should be taken into account when interpreting spaxel-scale $\sigma_{\rm gas}$ trends with line ratios.

Overall, we find no compelling evidence that shocks dominate the LINER-like excitation in this system, 
while explicitly acknowledging that the present diagnostics cannot exclude a localized or sub-dominant shock contribution. 
The global spaxel sample does not show a positive association between low-ionization line ratios and velocity dispersion, 
and the comparison to radiative shock model grids (Appendix~\ref{sec:shock_models}; Figure~\ref{fig:allen08_grid}) does not yield 
a unique or compelling match to the observed spaxel distribution without substantial parameter tuning. 
An independent check using the [S\,{\sc ii}]\,$\lambda\lambda$6717,6731 doublet ratio as a gas-pressure diagnostic (Appendix~\ref{sec:sii_ratio}) 
further supports this interpretation in the diffuse, 
low-density medium to which the [S\,{\sc ii}] doublet is sensitive: the low-$R$ ($R \equiv F(\lambda6717)/F(\lambda6731)$) spaxels are 
not associated with enhanced $\sigma_{\rm gas}$, as would be expected if radiative shocks dominated the excitation.

Three caveats apply to this kinematic test. First, the [O\,{\sc i}]/H$\alpha$ partial correlation at S/N\,$ \geq 5$ leaves room 
for a localized shock contribution in the partially ionized zone, 
as discussed above. Second, the [S\,{\sc ii}] doublet is collisionally saturated at $n_{\rm e} \gtrsim 10^3$--$10^4$\,cm$^{-3}$ 
and is intrinsically insensitive to compressed gas in radiative-shock cooling zones or in clumpy AGN-driven outflows \citep{Kakkad18, Baron19}, 
so the kinematic comparison constrains only the diffuse component. 
Third, our single-component Gaussian fits are limited by the spectral resolution of AAOmega ($R_{\rm blue}=1882$, $R_{\rm red}=4507$) and 
by the typical S/N of LINER-like spaxels in our data. 
A faint broad kinematic wing arising from a sub-dominant shock outflow could, in principle, 
lie below the detection threshold of our fits; 
a rigorous multi-component decomposition and quantification of the corresponding detection limits, 
following analyses applied to deeper IFS data \citep{Ho14, Davies14, Schreiber18}, 
require deeper observations than the current Hector data afford for this target and 
will be pursued in a forthcoming statistical study of the full Hector sample. 
Direct probes of the high-density gas (e.g., [Ar\,IV]\,$\lambda\lambda4711,4740$, 
which is not detected even in our central, highest-S/N spectrum) would similarly be required to test the saturated regime directly.

We further note that the AAOmega blue arm covers the auroral [O\,{\sc iii}]\,$\lambda4363$ line at the cluster redshift 
(observed at approximately 4604\,\AA), 
but no statistically significant emission is detected at this wavelength even in the highest-S/N central spectrum, 
where the strong H$\beta$ emission in the same spectral window confirms that the non-detection is intrinsic 
rather than an artifact of wavelength coverage or sensitivity. 
The lack of a direct electron-temperature constraint from [O\,{\sc iii}]\,$\lambda4363/\lambda5007$ means that 
our data cannot independently test the shock-heated, 
$T_e > 20{,}000$\,K scenario recently inferred for extended low-ionization emission 
from massive IFU stacking analyses of quiescent galaxies \citep{Lee24}. 
The presence of a dominant or localized shock-heating contribution at temperatures above the standard Case~B regime 
therefore cannot be conclusively excluded for this target on the basis of the present optical data alone.

The high-EW subset is deliberately biased toward spaxels where shocks, if 
present, would be most detectable, yet it comprises only $N = 33$ spaxels 
(about one third of the full set of spaxels used in this analysis), so any 
trends seen in this subset should not be interpreted as representative of 
the galaxy-wide excitation. We next turn to the Chandra X-ray data to test 
whether nuclear accretion can account for the required ionization.

\begin{deluxetable*}{ccccccccc}
\tabletypesize{\small}
\tablecaption{X-ray Spectral Fitting Results for the Hector LINER}
\label{tab:xrayfit}
\tablewidth{0pt}
\tablehead{
\colhead{ObsID} &
\colhead{Redshift} &
\colhead{$\log N_{\rm H}$} &
\colhead{$\Gamma$} &
\colhead{$\log L_{\rm 2-10}^{\rm obs}$} &
\colhead{$\log L_{\rm 2-10}^{\rm unabs}$} &
\colhead{Count Rate} &
\colhead{Hardness Ratio} &
\colhead{$C$ (dof)} \\
\colhead{(1)} & \colhead{(2)} & \colhead{(3)} & \colhead{(4)} & \colhead{(5)} & \colhead{(6)} & \colhead{(7)} & \colhead{(8)} & \colhead{(9)}
}
\startdata
5753 &
0.0553 &
$21.84^{+0.21}_{-0.30}$ &
$1.80^{+0.47}_{-0.42}$ &
$41.43^{+0.04}_{-0.07}$ &
$41.46^{+0.10}_{-0.10}$ &
$2.1\times10^{-3}$ &
$-0.0386$ &
$142.8\ (178)$ \\
\enddata
\tablecomments{
Column (1): Chandra observation ID.\\
Column (2): Adopted redshift (cluster redshift used for the luminosity calculation).\\
Column (3): Intrinsic hydrogen column density $\log N_{\rm H}$ (cm$^{-2}$) from XSPEC fit with \texttt{phabs*zphabs*powerlaw}; 
uncertainties are 90\% confidence intervals.\\
Column (4): Photon index $\Gamma$ of the power-law model with 90\% uncertainties.\\
Column (5): Observed (absorbed) rest-frame $2$--$10$ keV luminosity $\log L_{\rm 2-10}^{\rm obs}$ (erg s$^{-1}$); 
uncertainties are taken from the low/high bounds in the XSPEC luminosity output.\\
Column (6): Absorption-corrected (fully unabsorbed; Galactic + intrinsic removed) rest-frame $2$--$10$ keV luminosity $\log L_{\rm 2-10}^{\rm unabs}$ (erg s$^{-1}$), 
computed using the \texttt{cflux}-based procedure; uncertainties correspond to the low/high bounds.\\
Column (7): Net source count rate (counts s$^{-1}$).\\
Column (8): Hardness ratio ${\rm HR}=(H-S)/(H+S)$ where $H$ and $S$ are counts in the hard ($2$--$8$ keV) and soft ($0.5$--$2$ keV) bands.\\
Column (9): Cash statistic and degrees of freedom from the XSPEC fit.}
\end{deluxetable*}

% ------------------------------------------------------------------------------------------------------------------------------------------------------------------
\begin{figure*}
\centering
\includegraphics[width=1.0\linewidth, angle=0]{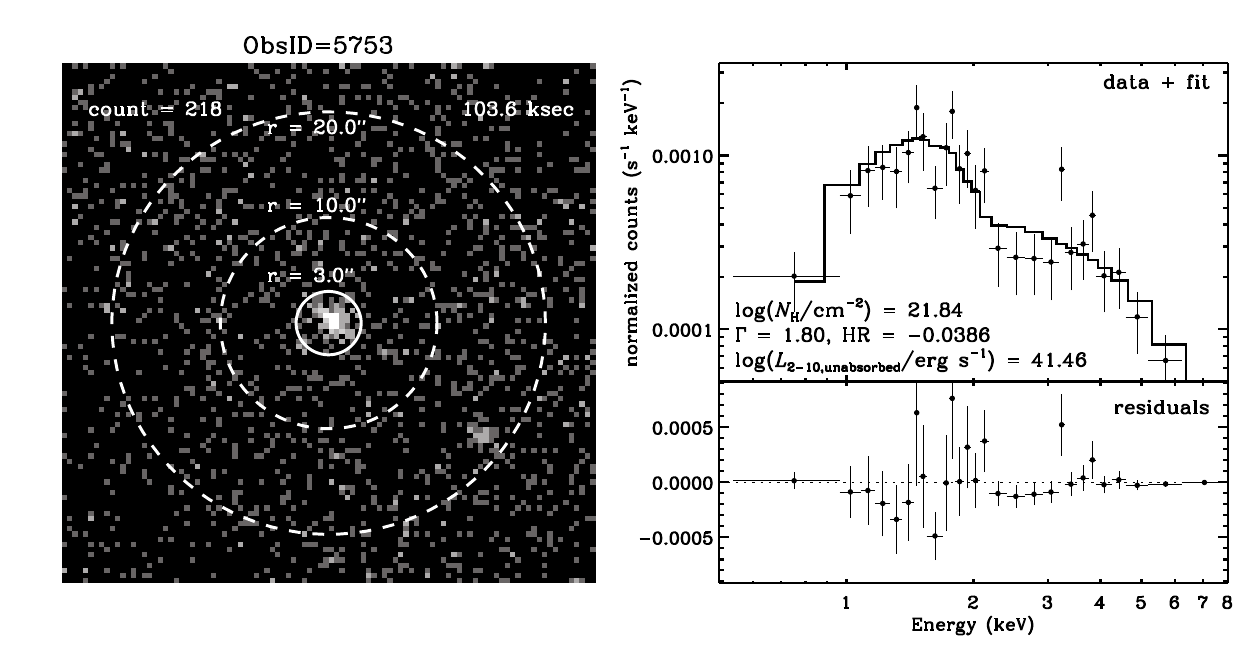}
\caption{Chandra X-ray data and model fit. 
Left panel: Chandra X-ray image, including source (solid circle) and background (dashed annulus) markers. 
Right panel: extracted data with the applied fit vs. energy (top) and residuals (bottom). 
\label{fig:xray}}
\end{figure*} % -------------------------------------------------------------------------------------------------------------------------------------------------

\subsection{Chandra X-ray observations and spectral fitting} \label{subsec:xray}
The Chandra X-ray observation of the galaxy (ObsID~5753) was obtained as part of deep imaging of the Abell~3667 cluster field, 
and was conducted on 2005 June 17 with a total exposure time of 103.63~ks.
We reprocessed the archival data using CIAO v4.17\footnote{\url{https://cxc.cfa.harvard.edu/ciao/}} and 
performed spectral analysis with XSPEC v12.14.1\footnote{\url{https://heasarc.gsfc.nasa.gov/docs/software/xspec/}} \citep{Arnaud96}.  
Our target is the only galaxy in our Hector LINER-like selection with overlapping archival Chandra coverage.
Accordingly, the present work is designed as a detailed case study rather than a statistical assessment of the LINER--AGN connection.

The source aperture, $A_{\rm s}$ (shown as a solid circle in Figure~\ref{fig:xray}), 
was defined based on the CIAO pipeline relation between off-axis angle and the recommended extraction radius, 
which accounts for the broadening of the Chandra PSF at larger off-axis angles. 
We also verified that the X-ray morphology is consistent with a point source by constructing a background-subtracted radial profile in the $0.5$--$7$\,keV band.
At the source off-axis angle ($\theta\simeq4.19^{\prime}$), we find that $\sim86\%$ of the net counts are enclosed within $2^{\prime\prime}$ and $\sim99\%$ within $5^{\prime\prime}$, 
with no compelling excess at larger radii; minor non-monotonicity at large radii is consistent with statistical over-subtraction in low-count outer annuli rather than extended emission.
A background annulus with a width of $10^{\prime\prime}$, centered on the source, was used to estimate the local background.
We extracted the spectrum using the \texttt{specextract} tool with the \texttt{psfcorr} option enabled, 
which applies an energy-dependent PSF correction to account for the fraction of source photons falling outside the extraction aperture.

A total of 218 net source counts were detected within the source aperture, 
after subtracting an expected background contribution of $\sim 20$ counts 
scaled from the background annulus (raw source-aperture counts: 238; raw background-annulus counts: 679; 
backscale ratio between the source aperture and the background annulus, $1:33.33$).
Fits were performed using the Cash statistic \citep{Cash79}, 
a maximum-likelihood estimator appropriate for low-count Poisson data, 
with the spectrum grouped to a minimum of 1 count per bin for numerical stability.
We modeled the spectrum in the rest-frame $0.5$--$8.0$\,keV band using an absorbed power-law model, \texttt{phabs*zphabs*powerlaw}, 
to account for both Galactic and intrinsic absorption, as well as the underlying AGN continuum emission.  

The best-fit photon index is $\Gamma = 1.80^{+0.47}_{-0.42}$,
with an intrinsic hydrogen column density of $\log (N_{\rm H}/{\rm cm}^{-2}) = 21.84^{+0.21}_{-0.30}$,
and an absorption-corrected $2$--$10$\,keV luminosity of
$\log L_{\rm X}^{\rm unabs} = 41.46^{+0.10}_{-0.10}$~\ergs.
For comparison, the observed (i.e., not absorption-corrected) $2$--$10$\,keV luminosity is
$\log L_{\rm X}^{\rm obs} = 41.43^{+0.04}_{-0.07}$~\ergs.
The hardness ratio is $-0.0386$, and the fit yields a Cash statistic of $C=142.8$ for 178 degrees of freedom.
The best-fit photon index $\Gamma = 1.80$ falls within the canonical range for AGN ($\Gamma \approx 1.5$--$2.0$), 
and in particular is in excellent agreement with the mean photon index of the Swift-BAT hard X-ray selected AGN sample 
($\langle\Gamma\rangle \approx 1.78$, $\sigma \approx 0.24$; \citealt{Ricci17}), 
providing independent spectral support for an accretion-powered origin of the nuclear X-ray emission. 
Here $\Gamma$ sets the slope of the intrinsic continuum,
while $N_{\rm H}$ preferentially attenuates soft X-ray photons through photoelectric absorption.
The resulting count spectrum reflects the combined effects of the intrinsic slope, absorption, and the instrument response.

As a sanity check, we also fitted the spectrum with a purely thermal plasma model (\texttt{phabs*apec}; 
two free parameters: $kT$ and normalization) 
and with a composite model (\texttt{phabs*(apec+powerlaw)}; 
three free parameters: $kT$, thermal normalization, and power-law normalization, with $\Gamma$ frozen at 1.8). 
Both models provide significantly worse fits than the absorbed power law. 
The Cash statistic increases by $\Delta C = +26.8$ for \texttt{phabs*apec} 
($\Delta\mathrm{dof} = +1$) and by $\Delta C = +18.9$ for \texttt{phabs*(apec+powerlaw)} 
($\Delta\mathrm{dof} = 0$; the power-law normalization is driven to zero). In both cases the thermal component 
converges to $kT = 64$~keV (the \textsc{xspec} hard limit), with the temperature unconstrained from above, 
indicating that the data do not require a thermal component at any physically meaningful temperature. 
We therefore adopt the absorbed power-law fit as our fiducial description of the nuclear X-ray emission.

A summary of the X-ray spectral modeling is presented in Table~\ref{tab:xrayfit}, and the data and best-fit model are shown in Figure~\ref{fig:xray}.

\subsection{Ionizing photon budget}
\label{subsec:budget}
To evaluate the dominant ionization mechanism responsible for the spatially
extended emission-line excitation, we compare the ionizing photon rate required
to reproduce the observed recombination-line luminosity with that available
from plausible ionizing sources.

We first estimate the Lyman-continuum photon rate required to power the observed \Ha\ emission.
The total extinction-corrected \Ha\ flux is obtained by summing over all spaxels with
${\rm S/N}_{\rm H\alpha} \geq 3$ (with no EW cut), because the spatially extended LINER-like emission
predominantly lies at low ${\rm EW}_{\rm H\alpha}$ and an EW threshold would systematically exclude a substantial
fraction of the diffuse component.

The extinction correction is performed on a spaxel-by-spaxel basis using the Balmer decrement.
For spaxels with ${\rm S/N}_{\rm H\alpha}\ge 3$ and ${\rm S/N}_{\rm H\beta}\ge 3$ and with an observed
$(\mathrm{H}\alpha/\mathrm{H}\beta) > 2.86$, we compute the color excess as
\begin{equation}
E(B\!-\!V)=\frac{2.5}{k_{\rm H\beta}-k_{\rm H\alpha}}
\log_{10}\left[\frac{(\mathrm{H}\alpha/\mathrm{H}\beta)}{2.86}\right],
\end{equation}
adopting $k_{\rm H\alpha}=2.535$ and $k_{\rm H\beta}=3.609$ from the Milky Way extinction curve of \cite{Cardelli89} with $R_{\rm V} = 3.1$.
The spaxel-level corrected H$\alpha$ flux is then
$F_{{\rm H\alpha},i}^{\rm corr}=F_{{\rm H\alpha},i}\,10^{0.4\,k_{\rm H\alpha}E(B\!-\!V)_i}$.
For ${\rm S/N}_{\rm H\alpha}\ge 3$ spaxels where $E(B\!-\!V)_i$ cannot be measured reliably (e.g., low-S/N H$\beta$ or
$(\mathrm{H}\alpha/\mathrm{H}\beta)\le 2.86$), we apply a single global $E(B\!-\!V)_{\rm glob}$ derived from the
Balmer decrement of the summed H$\alpha$ and H$\beta$ fluxes over the Balmer-decrement-valid spaxels, i.e.,
$F_{{\rm H\alpha},i}^{\rm corr}=F_{{\rm H\alpha},i}\,10^{0.4\,k_{\rm H\alpha}E(B\!-\!V)_{\rm glob}}$.
Summing the corrected fluxes over all spaxels with ${\rm S/N}_{\rm H\alpha} \geq 3$ yields a total extinction-corrected flux of
$F_{\rm H\alpha}^{\rm corr,tot}=4.10\times10^{-15}~{\rm erg~s^{-1}~cm^{-2}}$. 
We note that this zero-extinction assumption in the outer region is strictly conservative 
in the context of our hypothesis. If hidden dust is present in the outer spaxels, 
the true $Q_{\rm req}$ would be higher than the value adopted here, 
deepening the local $\tau$ deficit and further strengthening the conclusion that 
pAGB stars alone cannot account for the observed ionization. 
The robustness check using analytical $E(B\!-\!V)$ perturbations of $\pm 0.1$ and 
$\pm 0.2$~mag (Section~\ref{subsec:tau}) supports this interpretation: 
under reduced-dust perturbations the fraction of spaxels satisfying $\tau \ge 1$ remains at 1.4\%, 
identical to the fiducial case, confirming that the qualitative conclusion is insensitive to 
plausible dust uncertainties in the outer region. 
We note that the enhanced-dust perturbations ($E(B\!-\!V)$ increased by $+0.1$ and $+0.2$~mag) 
effectively test the scenario of hidden dust in the outer region, 
where the fiducial $E(B\!-\!V)_{\rm glob} \approx 0$ leaves the largest room for upward perturbation, 
and the fraction of spaxels with $\tau \ge 1$ drops to 0.0\% under both enhanced-dust cases, 
reinforcing the conservative nature of the original assumption.

At a luminosity distance of 246.8~Mpc, the corresponding extinction-corrected H$\alpha$ luminosity is
$L_{\rm H\alpha}^{\rm corr,tot}=2.99\times10^{40}~\mathrm{erg~s^{-1}}$.
A Monte Carlo propagation of the line-flux and Balmer-decrement uncertainties yields
a 16th--84th percentile range of $\sim0.03$~dex in $\log L_{\rm H\alpha}^{\rm corr,tot}$,
indicating that the uncertainty budget is dominated by the extinction correction.
Assuming Case~B recombination under typical low-density nebular conditions ($T_{\rm e}=10^{4}$~K),
the required ionizing photon rate is \citep{Osterbrock06}
\begin{equation}
Q_{\rm req}=7.3\times10^{11}
\left(\frac{L_{\rm H\alpha}^{\rm corr,tot}}{\mathrm{erg~s^{-1}}}\right)
\mathrm{photons~s^{-1}} ,
\label{eq:qreq}
\end{equation}
which yields
$Q_{\rm req}=2.18\times10^{52}~\mathrm{photons~s^{-1}}$,
or $\log Q_{\rm req}=52.34$.
This calculation assumes uniform absorption of ionizing photons,
neglecting possible clumpiness or geometric effects of the ionized gas that
could lead to photon escape or inefficiency in recombination.

The coefficient in Equation~(\ref{eq:qreq}) depends weakly on nebular conditions through the Case~B recombination coefficients. 
Adopting a lower electron temperature more appropriate for diffuse ionized gas (e.g., $T_{\rm e}=5000$--8000~K) 
decreases the coefficient from $7.3\times10^{11}$ to $\simeq(6.8$--$7.0)\times10^{11}$,
yielding $Q_{\rm req}\approx(2.03$--$2.09)\times10^{52}~{\rm s^{-1}}$ ($\log Q_{\rm req}\approx52.31$--52.32). 
This $\lesssim0.03$~dex shift does not affect our conclusions, 
which are dominated by uncertainties in the covering factor/escape fraction and by the spatially resolved $\tau$ deficit.

Stellar population synthesis models predict that pAGB stars and HOLMES, 
which dominate the ionizing budget in populations older than $\sim$100 Myr, 
produce a quasi-steady specific ionizing photon rate of 
$q_{\rm pAGB} \sim 10^{40}$--$10^{41}$~photons~s$^{-1}$~$M_\odot^{-1}$, 
with model-to-model variation of up to $\sim$1 dex depending on the adopted 
stellar evolutionary tracks, stellar atmospheres, and initial mass function (IMF) \citep[e.g.,][]{Binette94, CidFernandes11, Byler19}. 
Adopting a stellar mass of $M_{\rm *}=10^{10.97}\,M_\odot$, as derived by M. Beom et al. (in prep.), 
the expected total ionizing photon output from pAGB stars is
\begin{equation}
Q_{\rm pAGB} \simeq (0.93\text{--}9.3)\times10^{51}~{\rm photons~s^{-1}},
\end{equation}
or equivalently, $\log Q_{\rm pAGB}\simeq 50.97\text{--}51.97$. 
Because the measured \Ha\ emission is concentrated within the central region ($\lesssim R_{\rm e}$), 
using the galaxy-wide stellar mass likely overestimates the local pAGB photon budget.  
We therefore interpret $Q_{\rm pAGB}$ above as a conservative upper limit 
(see Section~\ref{subsec:tau} for further discussion). 
Our spatially resolved $\tau$ analysis partially mitigates this by distributing $M_\star$ across spaxels using a continuum proxy, 
thereby scaling the pAGB budget to the footprint of the emission. 
This estimate further assumes a high effective covering factor, such that most of the stellar Lyman-continuum photons are absorbed by surrounding warm gas 
(see, e.g., \citealt{Belfiore16}).

Finally, we estimate the ionizing photon output from the AGN based on the unabsorbed
$2$--$10$\,keV X-ray luminosity from the Chandra spectral analysis (Section~\ref{subsec:xray}), 
$\log L_{\rm X}=41.46$~erg~s$^{-1}$.
LLAGN with radiatively inefficient accretion flows often have relatively small bolometric corrections.
We adopt $k_{\rm bol}\sim10$--20 (e.g., \citealt{Ho08, Eracleous10}).
For comparison, $k_{\rm bol}\sim20$ is often adopted for Swift-BAT AGNs in the BAT AGN Spectroscopic Survey%
\footnote{\url{https://bass-survey.com}} \citep{Koss17, Koss22_data},
so that $L_{\rm bol}=k_{\rm bol}L_{\rm X}$ (i.e., $\log L_{\rm bol}\simeq42.46$--42.76).

% ------------------------------------------------------------------------------------------------------------------------------------------------------------------
\begin{figure*}
\centering
\includegraphics[width=1.0\linewidth, angle=0]{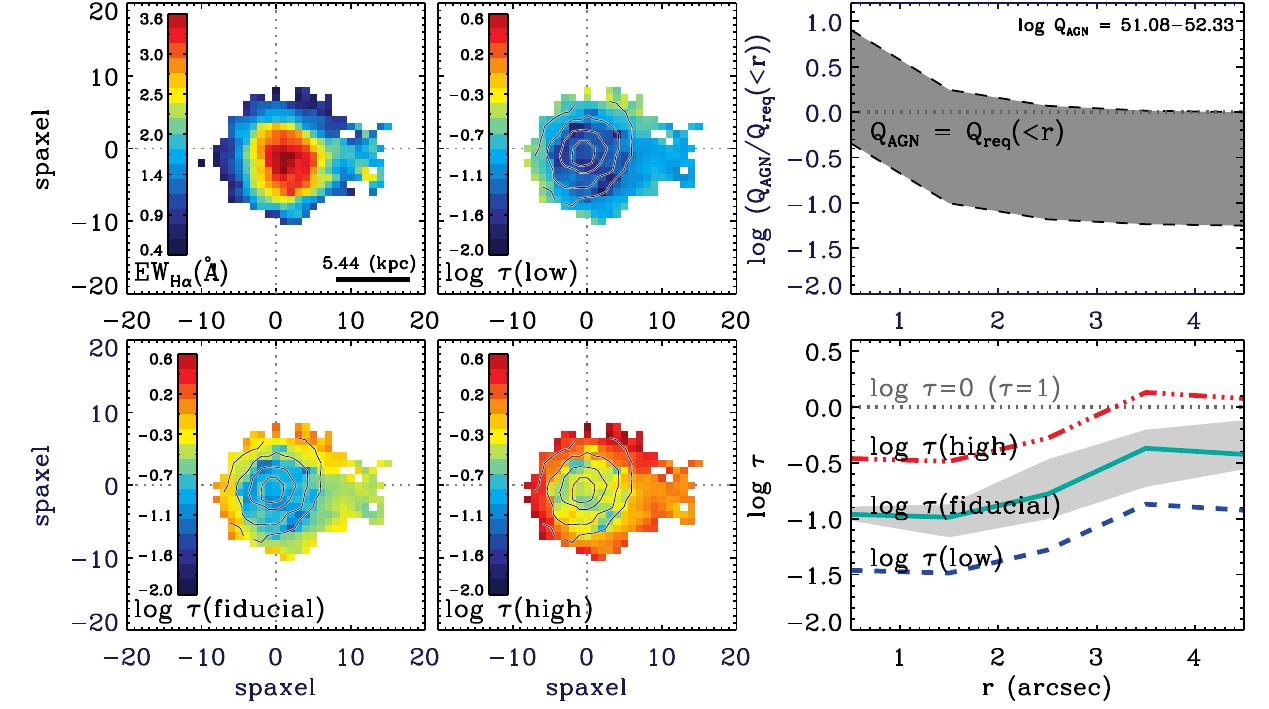}
\caption{Spatially resolved ${\rm EW}_{\rm H\alpha}$ and ionizing-photon budget diagnostics.
The top-left panel shows the ${\rm EW}_{\rm H\alpha}$ map.
The three map panels in the left and middle columns show $\log\tau \equiv \log(Q_{\rm pAGB}/Q_{\rm req})$ computed for each spaxel 
for three pAGB ionizing-photon normalizations, $q_{\rm pAGB}=10^{40.0}$, $10^{40.5}$, and $10^{41.0}$~photons~s$^{-1}$~$M_\odot^{-1}$ (low, fiducial, and high, respectively).
Black contours overlaid on the three $\log\tau$ map panels trace the stellar continuum proxy 
$f_{\rm cont}\equiv F({\rm H}\alpha)/{\rm EW}_{\rm H\alpha}$ used to define the local stellar-mass surface density 
$M_\star(i)$ in Section~\ref{subsec:tau}. Contour levels correspond to the 50th, 75th, 90th, and 97th percentiles 
of $\log_{10} f_{\rm cont}$ across the valid spaxels, marking the nucleus, the half-light region, 
and the outskirts of the stellar light distribution.
The top-right panel shows the cumulative AGN photon-budget ratio, 
$\log\!\left[Q_{\rm AGN}/Q_{\rm req}(<r)\right]$, 
where $Q_{\rm req}(<r)$ is the cumulative required ionizing-photon rate within projected radius $r$; 
the shaded region spans $\log Q_{\rm AGN}=51.08$–52.33.
The upper bound requires extreme bolometric corrections and ionizing fractions (see Section~\ref{subsec:budget}). 
The bottom-right panel presents the radial profile of $\log\tau$, computed as annular medians; the shaded band marks the 16th–84th percentile range for the fiducial case.
The horizontal dotted line at $\log\tau=0$ ($\tau=1$) indicates where the pAGB photon output matches the ionizing requirement implied by the observed H$\alpha$ emission.
Radial statistics include only bins containing at least $N_{\rm spaxel}\ge20$ spaxels with ${\rm S/N}_{\rm H\alpha}\ge3$; bins below this threshold are omitted.
\label{fig:tau}}
\end{figure*} % -------------------------------------------------------------------------------------------------------------------------------------------------

Assuming that a fraction $f_{\rm ion}$ of the bolometric luminosity is emitted
intrinsically in the ionizing continuum, 
the AGN ionizing photon rate can be written as
\begin{equation}
Q_{\rm AGN}=\frac{f_{\rm ion}\,L_{\rm bol}}{\langle h\nu\rangle},
\end{equation}
where $\langle h\nu\rangle$ is the characteristic energy per ionizing photon.
We further adopt $f_{\rm ion}\sim0.10$--0.30, motivated by typical AGN spectral energy distribution (SED) shapes (e.g., \citealt{Elvis94, Richards06}),
and a representative ionizing-photon energy $\langle h\nu\rangle\sim50$--150~{\rm eV}.
This yields
\begin{equation}
Q_{\rm AGN}\sim(1.20\times10^{51}\text{--}2.16\times10^{52})~\mathrm{photons~s^{-1}},
\end{equation}
or equivalently,
\begin{equation}
\log Q_{\rm AGN}\simeq51.08\text{--}52.33.
\end{equation}

We emphasize that $Q_{\rm AGN}$ estimated above represents the intrinsic ionizing-photon production of the AGN, 
and does not necessarily imply that the same photon rate is available to ionize the circumnuclear and extended gas. 
In particular, the nucleus is optically type~2, 
so the ionizing/UV continuum is likely obscured along our line of sight and may escape anisotropically 
(e.g., through a biconical opening). 
The measured column density ($\log N_{\rm H}\simeq21.8$) suggests strong attenuation of the ionizing/UV continuum 
for a Galactic gas-to-dust ratio, although AGN absorbers can be dust-poor.
The effective photon supply to the warm ionized gas therefore depends on the geometry, 
escape fraction, and covering factors of both the obscurer and the line-emitting gas. 
Our photon-budget comparison should thus be interpreted as an energetic upper bound: 
matching $Q_{\rm req}$ requires that a non-negligible fraction of the intrinsic ionizing output escapes the obscuring material, 
and is intercepted by the warm ionized gas.

The intrinsic $Q_{\rm AGN}$ range approaches the required photon rate ($\log Q_{\rm req}=52.34$)
and can match it only for parameters near the extreme high-$k_{\rm bol}$, high-$f_{\rm ion}$, low-$\langle h\nu\rangle$ end 
(or if the mean ionizing-photon energy is slightly below 50~eV or the bolometric correction slightly above 20).
It is also comparable, in order of magnitude, to the maximum ionizing output expected from pAGB stars
($\log Q_{\rm pAGB}\simeq 50.97\text{--}51.97$) under the assumption of a high effective covering factor.

Therefore, the global photon-budget comparison provides an order-of-magnitude consistency check 
rather than a unique decomposition of the ionizing sources.
Given the uncertainties in $q_{\rm pAGB}$, $k_{\rm bol}$, and the effective covering factor, 
either pAGB stars or the AGN could, in principle, supply a substantial fraction of $Q_{\rm req}$ in an integrated sense.
Discriminating which source dominates therefore requires a spatially resolved test of the photon budget. 
We provide such a test using the $\tau$ diagnostic in Section~\ref{subsec:tau}.

\subsection{Testing pAGB photoionization with the \texorpdfstring{$\tau$}{tau} diagnostic}
\label{subsec:tau}

While the global photon-budget comparison in Section~\ref{subsec:budget} provides an energetic plausibility check, 
it does not address whether pAGB/HOLMES photoionization can account for the emission locally across the IFU field. 
We therefore compute the spatially resolved $\tau$ parameter, 
defined as $\tau \equiv Q_{\rm pAGB}/Q_{\rm req}$, and examine maps of $\log\tau$. 
The $\tau$ diagnostic provides an empirical photon-budget test that is closely related to full photoionization modeling (e.g., \textsc{Cloudy}; \citealt{Ferland17}), 
but is restricted to the ionizing-photon requirement implied by the \Ha\ recombination luminosity. 
This formulation is readily applicable to large IFS samples, 
whereas a full photoionization analysis incorporating the observed line ratios as additional constraints is beyond the scope of this work.

For each spaxel, $Q_{\rm req}$ is derived from the extinction-corrected \Ha\ luminosity assuming Case~B recombination, 
using the same conversion as in Equation~(\ref{eq:qreq}). 
To estimate $Q_{\rm pAGB}$ per spaxel, we distribute the total stellar mass according to a continuum proxy 
such that the inferred stellar-mass surface density follows the observed stellar light and then 
adopt a specific pAGB ionizing-photon rate per unit stellar mass $q_{\rm pAGB}$ (in photons~s$^{-1}$~$M_\odot^{-1}$).

In practice, the local stellar continuum is estimated from ${\rm EW}_{\rm H\alpha}$ 
using the relation $f_{\lambda,{\rm cont}}(i) \simeq F_{{\rm H}\alpha}(i)/{\rm EW}_{{\rm H\alpha}}(i)$ for spaxels with ${\rm S/N}_{\rm H\alpha}\ge 3$. 
Each spaxel is then assigned a stellar mass by normalizing this proxy to the total stellar mass: 
$M_\star(i)=\left[f_{\lambda,{\rm cont}}(i)/\sum
f_{\lambda,{\rm cont}}\right]\,M_{\star,{\rm tot}}$, 
so that the inferred stellar-mass surface density traces the observed optical light to first order. 
We treat the resulting $M_\star(i)$ and $\tau(i)$ values as approximate, 
and focus on the radial behavior of the fiducial annular-median $\log\tau(r)$ rather than the exact value in any individual spaxel.

Given the $\sim$1 dex uncertainty in $q_{\rm pAGB}$, we compute $\log\tau$ for three representative normalizations, 
$q_{\rm pAGB}=10^{40.0}$, $10^{40.5}$, and
$10^{41.0}$ photons~s$^{-1}$~$M_\odot^{-1}$, referred to as the low, fiducial, and high cases, respectively. 
In this framework, $\log\tau=0$ ($\tau=1$) indicates that the pAGB photon budget matches the local ionizing requirement, 
whereas $\log\tau<0$ indicates a photon deficit.

Figure~\ref{fig:tau} shows the spatially resolved ${\rm EW}_{\rm H\alpha}$ and $\log\tau$ maps for the low, fiducial, and high pAGB normalizations, 
together with radial profiles. The ${\rm EW}_{\rm H\alpha}$ map (top-left) provides contextual information for the continuum proxy used to assign $M_\star(i)$, 
highlighting that the extended emission is dominated by low equivalent widths. The top-right panel shows the cumulative AGN photon-budget ratio, 
$\log\!\left(Q_{\rm AGN}/Q_{\rm req}(<r)\right)$, where $Q_{\rm req}(<r)$ is the cumulative recombination-required ionizing-photon rate within 
projected radius $r$. In the bottom-right panel, the $\log\tau(r)$ profiles are computed as annular medians, 
and for the fiducial case we additionally show the 16th--84th percentile range to indicate the spaxel-to-spaxel scatter at fixed radius. 

The innermost point corresponds to the median within a nuclear aperture ($r<1^{\prime\prime}$), 
and the remaining points are computed in annuli of width $\Delta r=1^{\prime\prime}$. 
We retain only bins containing at least $N_{\rm spaxel}\ge20$ spaxels with ${\rm S/N}_{\rm H\alpha}\ge3$. 
With a spaxel scale of $0.5^{\prime\prime}$, this criterion yields profiles over $r\simeq0.5$--$4.5^{\prime\prime}$. 

Over the radial range with robust spaxel coverage, the fiducial annular median $\log\tau$ remains below zero, 
showing that pAGB photoionization alone fails to close the local photon budget. 
Within the plotted footprint ($N=281$ spaxels with ${\rm EW}_{\rm H\alpha}>0$ and defined $\log\tau$), 
$1.4\%$ of spaxels satisfy $\log\tau\ge0$ in the fiducial case, with an additional $3.2\%$ in the range $-0.1\le\log\tau<0$. 
Under the high-$q_{\rm pAGB}$ normalization, these fractions increase to $39.9\%$ ($\log\tau\ge0$) and $17.8\%$ ($-0.1\le\log\tau<0$), respectively. 
Because the radial trends trace annular medians, 
a small fraction of high-$\tau$ spaxels visible in the maps does not necessarily shift the annular medians to $\log\tau\ge0$. 
Only under the high-$q_{\rm pAGB}$ normalization does the annular median $\log\tau$ approach zero at intermediate radii, 
whereas the nuclear region remains at $\tau<1$. 
Even under the optimistic normalization, the nuclear aperture therefore yields $\tau<1$, 
indicating that evolved stellar populations alone cannot account for the required ionization in the nucleus. 
This shortfall suggests an additional ionizing source, such as an AGN.

Several systematic effects should be noted when interpreting the $\tau$ maps. 
Algebraically, $F_{\mathrm{H}\alpha,\mathrm{obs}}$ appears in both $M_*(i)$ (via the continuum proxy) and $Q_{\mathrm{req}}(i)$ and cancels in the ratio, 
so that $\tau(i)$ reduces to a global prefactor divided by ${\rm EW}_{\rm H\alpha}(i)\times 10^{0.4\,k_{\mathrm{H}\alpha}\,E(B\!-\!V)_i}$. 
At fixed extinction, $\log\tau$ is therefore a linear transformation of $\log\,{\rm EW}_{\rm H\alpha}$ with slope~$-1$, 
and the anti-correlation between the $\tau$~and ${\rm EW}_{\rm H\alpha}$ maps in Figure~\ref{fig:tau} is expected by construction. 
The spatially varying extinction correction breaks this exact proportionality, 
and the physical advantage of $\tau$ over a single empirical ${\rm EW}_{\rm H\alpha}$ boundary (e.g., 3~\AA; \citealt{CidFernandes11}) is that 
it incorporates the actual stellar mass, luminosity distance, and spaxel-level extinction into a calibrated photon-budget test.

We note that $M_{\star,{\rm tot}}$ adopted here is the galaxy-wide stellar mass estimate, and is not restricted to the Hector IFU field of view. 
Our procedure therefore redistributes $M_{\star,{\rm tot}}$ across only those spaxels where the continuum proxy is defined, 
while preserving $\sum_i M_\star(i)=M_{\star,{\rm tot}}$ by construction. 
If a non-negligible fraction of the stellar mass lies outside the IFU footprint, 
this redistribution can overestimate the local stellar mass (and hence $Q_{\rm pAGB}$) within the mapped region. 
In that sense, the resulting $\tau$ values should be interpreted conservatively, 
because using a galaxy-wide $M_{\star,{\rm tot}}$ can bias $\tau$ high, and thus acts against concluding that an additional ionizing source is required.

We do not explicitly model spatial variations in the stellar mass-to-light ratio across the field. 
To anchor the magnitude of this systematic empirically, 
we note that spatially resolved spectroscopic studies of early-type and quiescent galaxies consistently find that 
the equivalent width of recombination lines (a tracer of the ionizing-photon-to-continuum ratio) fluctuates by 
$\simeq 32\%$ ($\simeq 0.12$~dex) around the mean value within individual galaxies \citep{Sarzi10}, 
and that radial stellar-population gradients exist in extended LIER galaxies 
but remain shallow over the regions probed by IFS \citep{Belfiore16}. 
The factor-of-two M/L systematic adopted here (corresponding to $\pm 0.30$~dex in $\log\tau$) therefore 
conservatively brackets the empirical scatter in the ionizing-photon-to-continuum ratio 
observed in comparable IFU samples. 
In the case of an approximately uniform M/L offset, 
$M_\star(i)$ and hence $\tau$ would be rescaled by the same factor. 
If M/L varies systematically with radius, it could introduce a mild radial tilt in the inferred $\tau(r)$ profile 
within the empirically constrained range. We treat such potential M/L gradients as a systematic uncertainty bounded 
by the literature range above, and focus on the robust qualitative result that 
the fiducial annular-median $\log\tau(r)$ remains below zero over the radial range with sufficient spaxel coverage.

As a further robustness check on the role of dust-extinction uncertainties, 
we recomputed $\log\tau$ after uniformly perturbing the spaxel-level $E(B-V)$ by $\pm 0.1$ and $\pm 0.2$~mag. 
These perturbation levels are designed to cover the systematic effects on the Balmer decrement arising from stellar continuum template choice, 
which can shift $A_V$ by $\sim$0.1~mag even in well-detected emission-line samples \citep{Groves12} 
and can produce $>$0.2~dex discrepancies in emission-line fluxes at low equivalent widths in IFS data \citep{Belfiore19}. 
The fraction of spaxels satisfying $\tau \geq 1$ remains at 1.4\% under both reduced-dust perturbations and is 0.0\% under both enhanced-dust perturbations, 
and the fiducial annular-median $\log\tau$ remains below zero at all radii. 
At larger radii ($r \gtrsim 3^{\prime\prime}$), the measured $E(B-V)$ values are already close to zero, 
so that reducing the extinction further has no effect on $\tau$. 
The qualitative conclusion is therefore insensitive to realistic dust uncertainties.

We note that valid spaxel-level $E(B-V)$ measurements from the Balmer decrement are confined to the inner $\sim$2.5$^{\prime\prime}$ ($\sim$2.7~kpc), 
where the values are highly patchy and show no coherent spatial structure. 
Beyond this region, the \Hb\ line is either too faint for a reliable Balmer decrement or the observed \Ha/\Hb\ ratio falls below the Case~B expectation, 
and the extinction correction relies on a single galaxy-averaged $E(B-V)$ estimate ($\approx 0$), 
confirming that the spatial morphology of the $\tau$ maps at larger radii is driven primarily by the ${\rm EW}_{\rm H\alpha}$ distribution 
rather than by spatially varying dust extinction.

The Hector PSF (FWHM\,$=$\,$2.75^{\prime\prime}$, corresponding to $\approx 3.0$~kpc) is comparable to 
the innermost radial bins, so beam smearing can redistribute centrally peaked emission into adjacent annuli. 
Because $\tau \propto 1/{\rm EW}_{\rm H\alpha}$ (at fixed extinction), 
this redistribution tends to dilute the nuclear ${\rm EW}_{\rm H\alpha}$ and hence inflate $\tau$ in the innermost bin, 
while depressing $\tau$ at $r\sim 1$--$2$ PSF~$\sigma$. 
The observed nuclear $\tau$ deficit is therefore a lower limit on the intrinsic deficit, 
reinforcing the conclusion that pAGB photoionization alone cannot close the photon budget in the nucleus. 
A PSF-deconvolved analysis is beyond the scope of this pilot study but would be valuable for larger samples.

\section{Discussion} \label{sec:discussion}

\subsection{Ionization budget and spatial test with the \texorpdfstring{$\tau$}{tau} diagnostic}
Our analysis shows that, although LINER-like emission is observed across nearly all 
spaxels with $\mathrm{S/N} \geq 3$ in the optical diagnostics,
identifying the dominant ionization mechanism requires a quantitative comparison of the available ionizing photon budgets.
Crucially, the availability of spatially resolved optical spectroscopy from the Hector IFU allows us
to compute the $\tau$ diagnostic on a spaxel-by-spaxel basis,
thereby constructing two-dimensional $\tau$ maps and radial profiles.
Rather than relying on a single aperture-integrated spectrum,
we can summarize $\tau$ in multiple annular apertures and directly track how any ionizing-photon deficit varies
with radius across the inner region of the galaxy.
The ionizing photon rate required to reproduce the extinction-corrected \Ha\ luminosity is $\log Q_{\rm req} = 52.34$.

For a stellar mass of $M_{\rm *}=10^{10.97}\,M_\odot$, the maximum ionizing photon output expected from pAGB stars and
HOLMES is $\log Q_{\rm pAGB} \simeq 50.97$--$51.97$, depending on the adopted specific ionizing-photon rate. 
The ionizing photon output inferred from the AGN's X-ray luminosity, $\log Q_{\rm AGN}\sim51.08$--$52.33$, 
spans a similarly broad range. 
At their respective upper limits both sources individually approach $Q_{\rm req}$, 
so neither can be excluded on purely energetic grounds. 
Conversely, at the lower normalizations both fall more than one dex short, 
underscoring that an integrated budget comparison alone cannot determine which source dominates. 
Distinguishing their relative contributions therefore demands the spatially resolved approach developed in Section~\ref{subsec:tau}. 
The spatially resolved $\tau$ measurements reveal that the ionizing photon deficit is not uniform, 
but is most pronounced in the inner region, where the contribution from a compact AGN is naturally expected to be strongest. 
As noted in Section~\ref{subsec:tau}, using the galaxy-wide $M_{\star,{\rm tot}}$ biases $\tau$ conservatively high, 
so the persistence of $\log\tau < 0$ cannot be attributed to an underestimated pAGB budget. 
Furthermore, PSF convolution tends to dilute centrally peaked emission into adjacent annuli, 
meaning that the observed nuclear $\tau$ deficit is a lower limit on the intrinsic deficit (Section~\ref{subsec:tau}).

\subsection{Assessing shocks, covering factor, and environmental effects}\label{subsec:assessing}

As an independent consistency check, we tested the shock-dominated scenario 
by comparing spaxel-scale low-ionization line ratios 
(\NIIHa, \OIHa, and \SIIHa) with the ionized-gas velocity 
dispersion $\sigma_{\rm gas}$. We find no robust positive correlation in the 
full spaxel set, and any suggestive trend in the high-EW subset becomes 
statistically inconclusive after controlling for projected radius 
(Figure~\ref{fig:shock_check}), indicating that radiative shocks are unlikely 
to dominate the extended LINER-like excitation (Section~\ref{subsec:shock}). 
We note, however, that this kinematic test constrains only the diffuse, 
low-density component to which the [S\,{\sc ii}] doublet is sensitive, and 
that the \OIHa\ partial correlation at S/N\,$ \geq 5$ leaves room 
for a localized, sub-dominant shock contribution in the partially ionized 
zone (see Section~\ref{subsec:shock} and Appendix~\ref{sec:sii_ratio} for details).

Independent of the shock assessment, the photon-budget estimates in Section~\ref{subsec:tau} assume a high effective covering factor for the warm ionized gas. 
If a significant fraction of ionizing photons escape or fail to be absorbed by the line-emitting gas, 
the actual photon budget required to sustain the observed emission would be larger, 
tightening constraints on the viability of evolved stars as the sole ionizing source.

As noted in Section~\ref{subsec:optical}, the target shows an asymmetric emission-line morphology 
consistent with RPS in the Abell~3667 environment (\c{C}ak{\i}r et~al., submitted). 
In cluster environments, RPS and related ICM--ISM interactions can generate extended, non-star-forming ionized gas and 
contribute to LINER-like excitation through heating and mixing in the multiphase medium (e.g., \citealt{Campitiello21}; \citealt{Owers19}; \citealt{Cakir26}). 
Importantly, an RPS-driven tail does not preclude an AGN. 
Jellyfish/RPS-affected systems can show elevated AGN incidence, 
plausibly linked to gas inflows triggered by stripping (e.g., \citealt{Poggianti17, Peluso22, KurinchiVendham25}).

Finally, we emphasize that the ``shock'' channel considered in our $\sigma_{\rm gas}$ tests is not restricted to star-formation feedback. 
Internal dynamical drivers may also generate localized shocks (e.g., bar-driven orbit crowding or AGN-driven outflows). 
The absence of a robust low-ionization line-ratio--$\sigma_{\rm gas}$ trend after controlling for projected radius suggests that such shocks, 
if present, are unlikely to be energetically dominant in powering the extended LINER-like component in this system. 
While our spatially resolved $\tau$ analysis and the compact nuclear X-ray source favor a substantial AGN contribution in the inner region, 
an additional environmental contribution to the most diffuse, large-radius component cannot be ruled out with the present data. 
A qualitative comparison to radiative-shock model grids is provided in Appendix~\ref{sec:shock_models}.

\subsection{Evidence for an LLAGN and accretion state from X-ray and MIR diagnostics}
\label{subsec:evidence}
The detection of a compact X-ray source in the Chandra data provides direct evidence for an AGN.
The spatial coincidence between this compact X-ray source and the region exhibiting the strongest $\tau$ deficit
provides a self-consistent picture in which nuclear accretion supplies a substantial fraction of the ionizing photons
required to power the extended LINER-like emission.
The unabsorbed $2$--$10$\,keV luminosity ($\log L_{\rm X}=41.46$~erg~s$^{-1}$) and photon index
($\Gamma=1.80$; Table~1) are consistent with those commonly observed in LLAGN.
Adopting bolometric corrections and ionizing fractions appropriate for radiatively inefficient accretion flows,
we estimate the ionizing photon rate from the AGN to be $\log Q_{\rm AGN}\sim51.08$--$52.33$.
This range overlaps with $Q_{\rm req}$ and supports the AGN as an energetically viable ionizing source.

The kpc-scale spatial extent of LINER-like line ratios in IFS data is often taken as evidence against a purely nuclear origin,
because classical narrow-line regions are typically compact.
In our target, however, ``AGN contribution'' does not require that the nucleus directly photoionizes every spaxel.
Instead, the spatially resolved $\tau$ analysis indicates that the photon deficit is most pronounced in the inner region,
coincident with the compact Chandra source, implying that an additional nuclear ionizing component is required at small radii.
At larger radii, the observed line emission can plausibly reflect a mixture of ionizing sources, where a low-ionization-parameter AGN continuum
illuminates diffuse, low-covering-factor gas while evolved stars contribute a quasi-extended baseline.
In addition, finite spatial resolution and PSF/beam-smearing effects can redistribute centrally peaked emission over several spaxels,
and the observed asymmetric morphology suggests that environmental processes (e.g., an RPS-related tail) may contribute to the most extended,
low-surface-brightness component.
Therefore, the extended LINER-like emission is most naturally interpreted as a composite of nuclear and extended ionizing channels,
with the nucleus required energetically in the inner region and additional diffuse contributions becoming increasingly important at larger radii.

Further support for an AGN origin of the ionizing photons, and specifically a radiatively inefficient accretion flow \citep{Narayan08},
comes from the host galaxy’s stellar velocity dispersion, measured within the effective radius to be
$\sigma_\star = 174 \pm 2~\mathrm{km~s^{-1}}$ (J. H. Lee et al., in prep.).
Using standard scaling relations between black hole mass ($M_{\rm BH}$) and stellar velocity dispersion ($\sigma_\star$; e.g., \citealt{Kormendy13}),
we infer $M_{\rm BH}\approx 10^{8.2}\,M_\odot$.
The resulting Eddington luminosity is $L_{\rm Edd}\approx 10^{46.3}~\mathrm{erg~s^{-1}}$.
Comparing this with the AGN bolometric luminosity inferred from the X-ray measurements
($L_{\rm bol}\approx 10^{42.4}\text{--}10^{42.7}~\mathrm{erg~s^{-1}}$) yields a low Eddington ratio of
$\log \lambda_{\rm Edd}\sim-3.9$ to $-3.6$.
This value firmly places the AGN in the regime of radiatively inefficient accretion \citep[e.g.,][]{Ho08, Yuan14}.
In such systems, the accretion likely proceeds via an advection-dominated accretion flow,
characterized by low radiative efficiency, weak mid-infrared emission,
and a reduced ionizing photon output compared to classical thin-disk AGNs.
This accretion mode is consistent with the observed multi-wavelength properties of the galaxy,
further supporting the interpretation of an LLAGN and
justifying the use of modest bolometric corrections and ionizing fractions in our photon budget analysis.

The object studied in this work does not fall within the AGN selection region in mid-infrared (MIR) diagnostics.
Its MIR colors, ${\rm W1}-{\rm W2}=0.11$ and ${\rm W2}-{\rm W3}=3.23$, derived from the
Wide-field Infrared Survey Explorer (WISE) all-sky survey \citep{Wright10},
lie well below commonly adopted AGN thresholds
(e.g., ${\rm W1}-{\rm W2} > 0.8$; \citealt{Stern12}; see also \citealt{Jarrett11, Mateos12}). 
The MIR non-detection does not independently confirm AGN activity, 
but is consistent with the expected properties of an LLAGN. 
The MIR and X-ray results are not in conflict, as MIR colors are primarily sensitive to hot-dust (torus) emission, 
which can be weak or absent in low-Eddington systems. This underscores the need for a multi-wavelength approach, 
with X-ray observations providing an important and complementary probe of weak nuclear activity that may not be captured by MIR color cuts.

\subsection{Evolved stellar photoionization and the role of AGN in extended LINER-like emission}
\label{subsec:pAGB_context}

A central question for interpreting extended LINER-like emission is whether evolved stellar populations alone can account for the observed ionization, 
or whether an additional source is required. 
\citet{Sarzi10} addressed this question systematically using SAURON integral-field data for a representative sample of nearby early-type galaxies. 
They found a remarkably tight correlation between the \Hb\ recombination line flux and the stellar surface brightness, 
holding not only in integrated measurements but also locally within individual galaxies, 
with the equivalent width of \Hb\ remaining nearly constant across the mapped fields. 
On the basis of this correlation and ionization balance calculations, 
they concluded that pAGB stars provide a sufficient Lyman continuum to power the diffuse nebular emission in the majority of their sample. 
Their analysis further demonstrated that AGN photoionization is consistent with the observed radial profiles 
only within the central 2--3~arcsec of the subset of galaxies hosting radio or X-ray cores, and that no AGN contribution is required to explain the extended emission. 
These findings were subsequently reinforced and extended to larger samples by \citet{Singh13} and \citet{Belfiore16}, 
who generalized the conclusion that spatially extended LINER-like emission is predominantly powered by evolved stellar populations.

Our spatially resolved $\tau$ analysis challenges the applicability of this framework to the present system. 
Under the fiducial pAGB normalization ($q_{\rm pAGB} = 10^{40.5}$~photons s$^{-1}$ $M_\odot^{-1}$), 
the annular-median $\log\tau$ remains below zero over the entire radial range with robust spaxel coverage, 
and only 1.4\% of spaxels satisfy $\tau \geq 1$ (Section~\ref{subsec:tau}). 
Even under the optimistic high normalization ($q_{\rm pAGB} = 10^{41.0}$), the nuclear aperture still yields $\tau < 1$. 
Combined with the absence of a robust shock signature (Section~\ref{subsec:shock}) and the Chandra detection of a coincident LLAGN (Section~\ref{subsec:xray}), 
these results identify the AGN as the most plausible source of the missing ionizing photons.

This finding is not at odds with the conclusions of \citet{Sarzi10}, 
who themselves noted that AGN can operate within the nuclear region of galaxies with established X-ray or radio cores. 
It extends their picture, however, by showing that the pAGB ionizing photon budget falls short not only in the nucleus, 
where the deficit is most pronounced and coincident with the compact X-ray source, but also across the broader mapped field. 
As discussed in Section~\ref{subsec:evidence}, the composite ionization scenario 
in which a dilute AGN radiation field supplements an evolved-stellar baseline at larger radii naturally accommodates this result 
without requiring that the nucleus directly photoionizes every spaxel. 
A key distinction from the \citet{Sarzi10} sample is that their tight \Hb--continuum correlation was established 
across a population of nearby early-type galaxies spanning a range of luminosities and morphologies, 
whereas the present study examines a single system in which independent multi-wavelength evidence confirms the presence of an LLAGN. 
The $\tau$ diagnostic provides a spaxel-level tool to test precisely where the pAGB budget succeeds or fails within such a system, 
going beyond the surface-brightness correlation approach.

The $\sim$1~dex uncertainty in $q_{\rm pAGB}$ prevents a definitive conclusion about the magnitude of the shortfall at each location, 
and the degree to which pAGB stars contribute remains normalization-dependent. 
Under the high normalization, $\sim$40\% of spaxels across the mapped region reach $\tau \geq 1$, 
and the annular-median $\log\tau$ approaches zero at intermediate radii, 
consistent with a scenario in which pAGB photoionization provides a substantial but incomplete baseline that is supplemented by AGN radiation. 
The key qualitative result is nonetheless robust to the choice of normalization. 
The fiducial and low cases both indicate a pervasive ionization deficit, and even the optimistic case cannot eliminate the nuclear shortfall. 
As noted in Section~\ref{subsec:tau}, this nuclear deficit is a lower limit on the intrinsic value 
because PSF convolution redistributes centrally peaked emission into adjacent annuli. 
This system thus illustrates that extended LINER-like emission cannot be automatically attributed to evolved stellar populations without a quantitative, 
spatially resolved assessment of the local ionizing photon budget, 
directly addressing the tension between pAGB-dominated and AGN-hosting interpretations of LINER-like galaxies raised in the Introduction.

\subsection{Comparison samples and broader implications}
\begin{figure*}
\centering
\includegraphics[width=1.0\linewidth, angle=0]{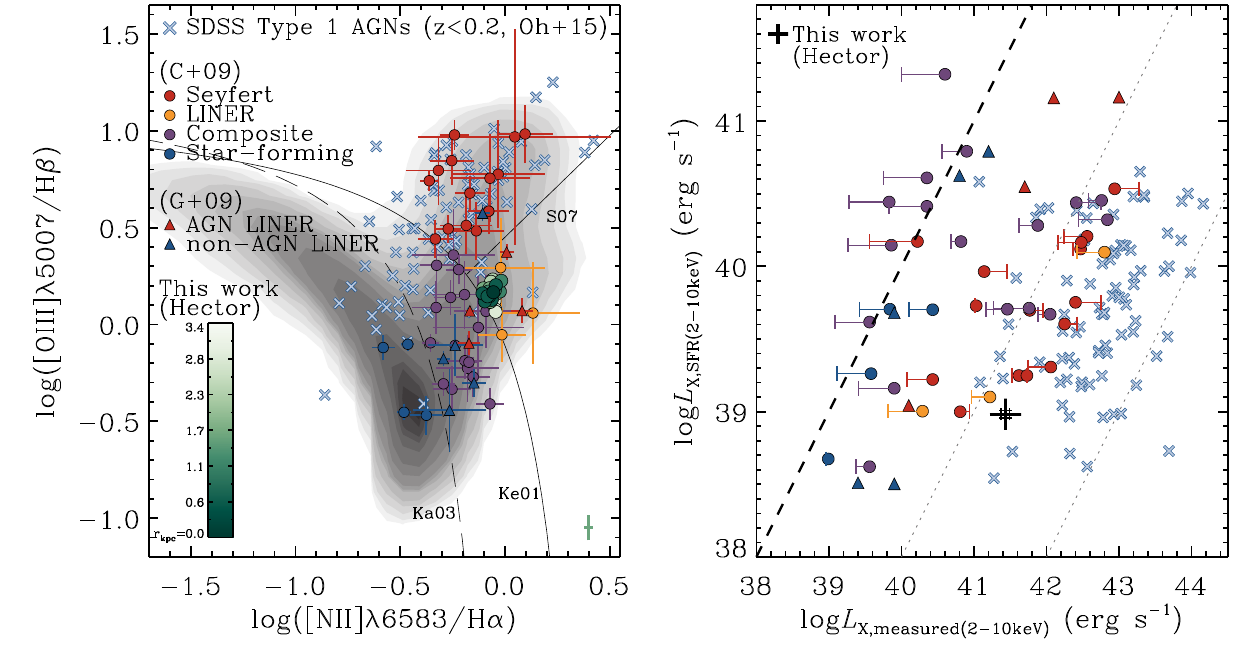}
\caption{Comparison with various classes of emission-line galaxies with X-ray detections. 
Left panel: \NIIHa\ BPT diagnostic diagram showing the position of our target (green filled circles, color-coded by projected distance from the galaxy center), 
compared to sources with different ionization mechanisms, including LINER-like galaxies and AGNs from the literature with X-ray detections 
\citep{Constantin09, GonzalezMartin09, Oh15}. 
Right panel: Comparison between the $2$--$10$\,keV luminosity inferred from the \Ha-based SFR,
$L_{\mathrm{X,SFR}}$, and the measured (spectral-fitting) value, $L_{\mathrm{X,measured}}$.
The thick dashed line marks the one-to-one relation,
$\log L_{\mathrm{X,SFR}}=\log L_{\mathrm{X,measured}}$.
The two dotted lines indicate offsets of $+2$ and $+4$ dex in $\log L_{\mathrm{X,measured}}$ relative to one-to-one,
i.e., $\log L_{\mathrm{X,SFR}}=\log L_{\mathrm{X,measured}}-2$ and
$\log L_{\mathrm{X,SFR}}=\log L_{\mathrm{X,measured}}-4$.
\label{fig:ranalli}}
\end{figure*}

Further support for an AGN origin of the ionizing source comes from comparisons with other LINER-like galaxies
exhibiting X-ray detections. In the left panel of Figure~\ref{fig:ranalli},
our target is shown on the \NIIHa\ BPT diagnostic diagram alongside a sample of emission-line galaxies,
including LINERs and AGNs, with confirmed X-ray emission.
We include a subset ($N=39$) of X-ray-detected galaxies from \citet{Constantin09},
for which all relevant emission lines are measured with high S/N ($ \geq 3$),
using classifications and line ratios from the OSSY catalog\footnote{\url{https://data.kasi.re.kr/vo/OSSY/}} \citep{Oh11}.
SDSS LINERs with X-ray detections compiled by \citet{GonzalezMartin09} are also shown,
with AGNs (red triangles, $N=4$) and non-AGNs (blue triangles, $N=5$) distinguished.
For comparison, we include a sample of SDSS type~1 AGNs
with broad Balmer lines and Chandra X-ray detections
(light-blue crosses, $N=79$) from \citet{Oh15}.

Despite spanning similar regions in the BPT diagnostic diagram,
these various classes illustrate that traditional optical diagnostic boundaries do not always reflect the true ionizing source,
particularly for LINER-like systems.
The right panel of Figure~\ref{fig:ranalli} compares unabsorbed $2$--$10$\,keV X-ray luminosities with those
expected from \Ha-based star formation rates for the same set of objects shown in the left panel,
where the SFR is inferred under the assumption that \Ha\ traces star formation (and therefore represents an upper limit).
Our object lies well below the one-to-one relation, i.e., the SFR-predicted $L_{\rm X}$ is much lower than the observed $L_{\rm X}$.
This discrepancy shows that X-ray binaries (XRBs) associated with star formation cannot account for the hard X-ray output,
implying an additional X-ray source, most naturally an LLAGN, as the dominant contributor to $L_{\rm X}$.

% ------------------------------------------------------------------------------------------------------------------------------------------------------------------
 % -------------------------------------------------------------------------------------------------------------------------------------------------

Although the compact Chandra source strongly suggests nuclear accretion,
a fraction of the observed $2$--$10$\,keV X-ray emission could in principle arise from XRBs within the extraction aperture.
We therefore estimate the expected $2$--$10$\,keV luminosity from XRBs
as an additional consistency check.
It can be approximated as the sum of contributions from low-mass and high-mass XRB populations
that scale with stellar mass and star-formation rate, respectively:
\begin{equation}
L_{\rm X,XRB}(2\!-\!10~{\rm keV})=\alpha\,M_\star+\beta\,{\rm SFR}.
\end{equation}
Adopting $\alpha = 9.05\times10^{28}$ erg s$^{-1}\,M_\odot^{-1}$ and
$\beta = 1.62\times10^{39}$ erg s$^{-1}\,(M_\odot~{\rm yr}^{-1})^{-1}$
\citep[e.g.,][]{Lehmer10}, and using a stellar mass of
$\log M_\star = 10.97$ ($M_\star = 9.33\times10^{10}\,M_\odot$) together with the
H$\alpha$-based star-formation rate
${\rm SFR}_{\rm H\alpha}=7.9\times10^{-42}L_{\rm H\alpha}\simeq0.25~M_\odot~{\rm yr}^{-1}$
for $L_{\rm H\alpha}=3.17\times10^{40}$ erg s$^{-1}$, we obtain
\begin{equation}
L_{\rm X,XRB}\approx8.85\times10^{39}~{\rm erg~s^{-1}}
\end{equation}
with the emission dominated by the low-mass XRB component.
As the spectrum is extracted from a nuclear aperture,
using the total galaxy stellar mass likely overestimates the stellar mass enclosed
within the extraction region and thus yields a conservative upper limit on $L_{\rm X,XRB}$.
This value is $\sim1.5$ dex below the observed unabsorbed nuclear luminosity
($\log L_{\rm X}=41.46$, in ${\rm erg\,s^{-1}}$), indicating that XRBs cannot account for the hard X-ray
output.

%  ------------------------------------------------------------------------------------------------------------------------------------------------------------------
\begin{figure*}
\centering
\includegraphics[width=0.86\linewidth, angle=0]{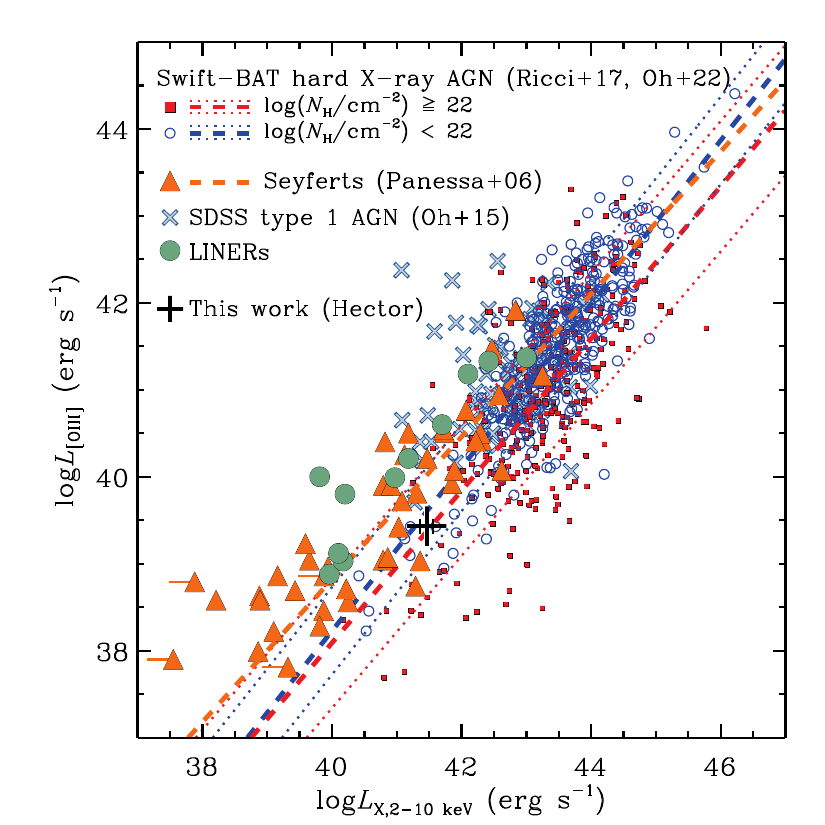}
\caption{Relationship between extinction-corrected \OIII\ luminosity and intrinsic (unabsorbed) $2$--$10$\,keV X-ray luminosity for various AGN classes.
Red-filled squares and blue open circles represent Swift-BAT hard X-ray selected AGNs with 
$\log(N_{\rm H}/{\rm cm^{-2}}) \ge 22$ (type 2) and $\log(N_{\rm H}/{\rm cm^{-2}}) < 22$ (type 1), respectively \citep{Ricci17, Oh22}.
Orange-filled triangles show Seyfert galaxies from \citet{Panessa06}.
Light-blue crosses indicate SDSS type 1 AGNs from \citet{Oh15}.
Green-filled circles denote LINERs compiled from the literature (see Table~\ref{tab:liner} for references). 
The black cross marks the Hector target presented in this work.
The orange dashed line shows the $\log L_{\mathrm{X}} - \log L_{\mathrm{[O\,III]}}$ relation from \citet{Panessa06}.
Blue and red dashed lines represent linear regression fits to the Swift-BAT type 1 and type 2 AGNs, respectively.
Dotted lines correspond to the $\pm 1\sigma$ range around each regression line.
\label{fig:LOIII_LX}}
\end{figure*} % -------------------------------------------------------------------------------------------------------------------------------------------------

In Figure~\ref{fig:LOIII_LX}, we explore the relationship between extinction-corrected \OIII\ luminosity
and unabsorbed $2$--$10$\,keV X-ray luminosity for different AGN populations.
The comparison sample includes Swift-BAT hard X-ray selected AGNs \citep{Koss17, Koss22_data},
which are divided into type~1 ($\log N_{\mathrm{H}} < 22~\mathrm{cm}^{-2}$, blue open circles)
and type~2 ($\log N_{\mathrm{H}} \ge 22~\mathrm{cm}^{-2}$, red filled squares) sources.
 
The Swift-BAT type~1 AGNs broadly follow the correlation reported by \citet{Panessa06},
whereas type~2 AGNs show a larger scatter and a systematic tendency toward lower \OIII\ luminosities at fixed $L_X$.
Quantifying offsets from the \citet{Panessa06} relation using residuals
$\Delta\log L_{\rm [OIII]}=\log L_{\rm [OIII]}-(0.82\,\log L_X+6.02)$,
Swift-BAT type~1 AGNs ($N_{\rm H}<10^{22}$ cm$^{-2}$; $N=330$) show a modest median offset of
$-0.16$ dex with an RMS scatter of 0.51 dex, whereas type~2 AGNs
($N_{\rm H}\ge10^{22}$ cm$^{-2}$; $N=323$) exhibit a larger negative median offset of
$-0.51$ dex and substantially increased scatter (RMS = 0.75 dex).
The stronger offset and scatter for type~2 AGNs are consistent with additional object-to-object diversity in
line-of-sight obscuration/complex absorption and orientation, as well as variations in narrow-line region
covering factor and extinction.

%----------Table --------------------------------------------------------------------------------------------------------------
\begin{deluxetable*}{lrrrcccc}[ht]
\tabletypesize{\small}
\tablecaption{Compilation of nearby LINERs with X-ray spectral detections}
\label{tab:liner}
\tablewidth{0pt}
\tablehead{
\colhead{Name} &
\colhead{R.A.} &
\colhead{Decl.} &
\colhead{Redshift} &
\colhead{$\log L_{\mathrm{[O\,III]}}$} &
\colhead{$\log L_{\mathrm{X}, 2-10 {\rm keV}}$ } &
\colhead{[O\,III] Ref.} &
\colhead{X-ray Ref.} \\
\colhead{(1)} & \colhead{(2)} & \colhead{(3)} & \colhead{(4)} & \colhead{(5)} & \colhead{(6)} & \colhead{(7)} & \colhead{(8)}
}
\startdata
UGC~05101              & 143.964992 & 61.353256 & 0.039367 & 41.18 & 42.10 & O11 & G09 \\
M81                        	& 148.888221 & 69.065295 & 0.000130 & 39.80 & 40.20 & G17 & P06 \\
NGC~3998                & 179.483889 & 55.453589 & 0.003401 & 40.21 & 41.18 & S15 & Y11 \\
NGC~4278                & 185.028439 & 29.280754 & 0.002165 & 38.88 & 39.96 & H97, B16 & T03 \\
NGC~4676B              & 191.546926 & 30.722734 & 0.021798 & 39.12 & 40.10 & O11 & G09 \\
NGC~5005                & 197.734437 & 37.059044 & 0.003156 & 39.03 & 40.17 & S22 & S22 \\
Mrk~266~NE              & 204.574048 & 48.278099 & 0.027699 & 40.60 & 41.70 & O11 & G09 \\
SDSS~J134054.60+400637.4    & 205.227732 & 40.110384 & 0.170775 & 41.33 & 42.42 & O11 & C09 \\
UGC~08696                  & 206.175463 & 55.886847 & 0.037340 & 41.37 & 43.00 & O11 & G09 \\
SDSS~J144242.62+011151.1    & 220.677617 &  1.319750 & 0.033674 & 39.99 & 40.97 & O11 & C09 \\
SDSS~J161740.54+350015.2    & 244.418909 & 35.004222 & 0.029830 & 40.00 & 39.81 & O11 & C09 \\
\enddata
\tablecomments{
All coordinates (J2000) and redshifts are taken from NED (\url{https://ned.ipac.caltech.edu}).
Column (5) gives extinction-corrected \OIII\ luminosities, and Column (6) shows unabsorbed $2$--$10$\,keV X-ray luminosities.
All luminosities are expressed in units of \ergs.
Columns (7) and (8) provide the references for $\log L_{\mathrm{[O\,III]}}$ and $\log L_{\mathrm{X}, 2-10 {\rm keV}}$, respectively.  
Reference codes: 
H97 = \citet{Ho97b}, 
T03 = \citet{Terashima03}, 
P06 = \citet{Panessa06}, 
C09 = \citet{Constantin09}, 
G09 = \citet{GonzalezMartin09}, 
O11 = \citet{Oh11}, 
Y11 = \citet{Younes11}, 
S15 = \citet{Saikia15}, 
B16 = \citet{Balmaverde16}, 
G17 = \citet{GomezGuijarro17}, 
S22 = \citet{Saade22}.
}
\end{deluxetable*}

Nearby LINERs with confirmed X-ray detections compiled from the literature (Table~\ref{tab:liner}) are also shown.
These objects, which are optically classified as LINERs but host AGNs based on their X-ray properties,
are distributed close to the locus of nearby AGNs studied by \citet{Panessa06}.
However, at fixed $L_{\rm X}$ they systematically lie above the Panessa best-fit relation,
indicating enhanced \OIII\ emission relative to nearby Seyferts at the same $L_{\rm X}$.
This offset suggests that, in many nearby LINERs, the observed \OIII\ luminosity may include a substantial contribution from
spatially extended ionized gas excited by mechanisms in addition to direct nuclear photoionization,
such as shocks or ionization by evolved stellar populations.
Ongoing star formation could also contribute locally,
but for our target the H$\alpha$-based SFR is an upper limit
and the corresponding XRB luminosity is $\sim$1.5 dex below
the observed $L_{\rm X}$, making star formation an unlikely
dominant explanation.
Alternatively, the offset may reflect significant X-ray absorption that suppresses the observed $2$--$10$\,keV luminosity.

In contrast, our Hector LINER target occupies a different location from many nearby LINERs in this diagram.
At fixed $\log L_X$, it lies 0.59 dex below the \citet{Panessa06} relation,
but is much closer to the Swift-BAT sequences, with offsets of $-0.17$ dex relative to the type~1 fit and
$+0.07$ dex relative to the type~2 fit.
This placement implies that the optical narrow-line emission is not strongly dominated by excess extended \OIII\ emission relative to the nuclear X-ray output,
and is instead broadly consistent with the nuclear X-ray source as the primary driver.
Such behavior is consistent with a compact narrow-line region
and a comparatively unobscured or mildly obscured view of the AGN,
in agreement with the X-ray spectral properties derived from the Chandra data.
 
Overall, this comparison highlights the diversity of LINER-like systems in the $\log L_{\mathrm{X}}$--$\log L_{\mathrm{[OIII]}}$ plane.
In particular, the \citet{Panessa06} $\log L_{\mathrm{X}}$--$\log L_{\mathrm{[OIII]}}$ scaling shown in Figure~\ref{fig:LOIII_LX},
which is calibrated on nearby Seyferts, is not a uniform descriptor of LINER-like galaxies.
Literature LINERs tend to lie above the relation at fixed $L_{\mathrm{X}}$, whereas our Hector target lies below it.
These trends suggest that the relative balance between optical and X-ray emission is sensitive to obscuration, geometry/orientation,
and the spatial extent (and contamination) of the narrow-line emitting gas.

\subsection{Prospects for larger IFS samples}\label{subsec:prospects}
 
This work combines a spatially resolved $\tau$-based ionizing photon-budget analysis
with an independent X-ray constraint on nuclear accretion within a single LINER-like galaxy.
The kpc-scale spatial information provided by the Hector IFU, together with deep Chandra observations,
allows us to disentangle the relative roles of nuclear and extended ionizing sources responsible for
the spatially extended LINER-like emission in a manner that is not accessible with single-aperture spectroscopy alone.
 
The same framework can be applied to a broader set of spatially extended LINER-like candidates in modern IFS surveys.
On the optical side, the practical requirements are reliable extinction-corrected recombination-line luminosities,
robust stellar-continuum modeling to estimate $Q_{\rm pAGB}$, and sufficient spatial sampling to construct $\tau \equiv Q_{\rm pAGB}/Q_{\rm req}$ maps and radial profiles.
Hector is particularly well suited to this because it will deliver homogeneous,
spatially resolved spectroscopy for a large number of nearby galaxies, enabling efficient identification of systems
with extended LINER-like line ratios and low ${\rm EW}_{\rm H\alpha}$ over kpc scales.
 
The key practical challenge in extending this approach is obtaining uniform, nuclear-specific X-ray constraints across a larger sample.
Because $Q_{\rm AGN}$ depends on the hard X-ray luminosity (and its absorption correction),
shallow or spatially coarse X-ray data can be limiting, especially in the presence of host-galaxy contamination.
A natural way forward is therefore a tiered strategy that leverages archival X-ray measurements where available
and obtains targeted follow-up for a subset selected to span the range of $\tau$ deficits and host-galaxy properties.
In this way, even a modest program that combines a large optical parent sample with X-ray constraints for a smaller,
well-defined subset can establish how common AGN-dominated ionization is among spatially extended LINER-like systems and 
provide empirical guidance for interpreting LINER-like excitation in future wide-area IFS surveys.

\section{Summary and Conclusions} \label{sec:conclusion}

We have investigated the physical origin of spatially extended LINER-like emission
in a low-redshift galaxy using optical IFS from the Hector Galaxy Survey together with archival Chandra data.
The goal of this work is not simply to re-identify LINER-like line ratios, but to determine
which ionizing source can plausibly power the emission once both global energetics and spatial variations are considered.
To this end, we combine a spatially resolved photon-budget test based on the $\tau$ diagnostic
with an independent nuclear constraint from X-ray spectral modeling.

Our optical diagnostics show LINER-like line ratios across nearly all spaxels with $\mathrm{S/N} \geq 3$,
and the spaxel-scale tests provide little support for radiative shocks as the dominant excitation mechanism.
Low-ionization line ratios (\NIIHa, \OIHa, \SIIHa) do not exhibit a robust correlation with $\sigma_{\rm gas}$
once projected radius is controlled for. 
We note, however, that the present diagnostics cannot exclude a localized or sub-dominant shock contribution, particularly within the partially ionized zone traced by \OI.
In terms of global energetics, the recombination-required ionizing photon rate
($\log Q_{\rm req}=52.34$) is approached by the upper range of pAGB expectations
($\log Q_{\rm pAGB}\sim 50.97$--$51.97$) and by the AGN ionizing output inferred from the X-ray luminosity
($\log Q_{\rm AGN}\sim 51.08$--$52.33$),
although at the lower normalizations both sources fall more than one dex short of $Q_{\rm req}$.
These estimates further assume that the warm ionized gas has a high effective covering factor
and that Lyman-continuum escape is not severe.
This motivates a spatial test, because if evolved stars dominate, the photon budget should close locally across the field.

Using $\tau \equiv Q_{\rm pAGB}/Q_{\rm req}$, we find that
for a fiducial normalization ($q_{\rm pAGB}=10^{40.5}~{\rm photons~s^{-1}}~M_\odot^{-1}$),
$\log\tau<0$ over most of the IFU footprint, indicating a systematic photon deficit relative to pAGB expectations.
Only under optimistic pAGB normalizations does $\tau$ approach unity at intermediate radii,
while the nuclear region remains consistently below unity.
These results imply that evolved stellar populations contribute to the ionization budget,
but cannot by themselves account for the observed emission across the bulk of the mapped region.
The $\tau$ estimates are conservative because $M_{\star,{\rm tot}}$ is redistributed over the IFU footprint (Section~\ref{subsec:tau}),
so the inferred $\log\tau < 0$ values would only become more negative if stellar mass outside the field were accounted for.
PSF convolution further implies that the observed nuclear deficit is a lower limit on the intrinsic value.

An AGN contribution is independently supported by the Chandra detection of a compact nuclear X-ray source.
The X-ray spectrum is consistent with an LLAGN, with $\Gamma \approx 1.8$ and an absorption-corrected luminosity
$\log L_{\rm X}(2$--$10~{\rm keV})=41.46~{\rm erg~s^{-1}}$.
The galaxy lies well above the star-forming $L_{\rm X}$--SFR expectation even when adopting the H$\alpha$-based SFR as a conservative upper limit,
and an explicit XRB estimate underpredicts the observed $L_{\rm X}$ by $\sim$1.5~dex,
ruling out stellar processes as the dominant origin of the nuclear hard X-ray emission.
The low Eddington ratio ($\log\lambda_{\rm Edd}\sim-3.9$ to $-3.6$) further supports radiatively inefficient accretion,
consistent with the weak mid-infrared AGN signatures.

Taken together, our results show that the extended LINER-like emission cannot be explained by pAGB photoionization alone
once spatially resolved photon budgets are applied.
A weak AGN is required at least in the inner region and likely contributes to the ionizing budget at larger radii as well,
though the precise radial extent of its contribution remains unconstrained.
Evolved stars contribute more substantially at larger radii.
An additional contribution from environmental processes (e.g., RPS in the Abell~3667 cluster)
to the most extended, low-surface-brightness component cannot be excluded with the present data.
This result provides a spatially resolved counterpoint to the prevailing interpretation
that extended LINER-like emission in early-type galaxies is predominantly powered by
pAGB photoionization \citep[e.g.,][]{Sarzi10, Singh13, Belfiore16},
demonstrating that a quantitative, spaxel-level photon-budget test
can reveal AGN contributions that are not apparent from
surface-brightness correlations or integrated diagnostics.
More broadly, this work shows that extended LINER-like emission can conceal a substantial LLAGN contribution
even when traditional optical and infrared AGN indicators are weak.
Applying the same $\tau$-plus-X-ray framework to the broader Hector survey,
complemented by targeted X-ray follow-up of systems spanning a range of $\tau$ deficits,
will establish the prevalence of LLAGN contributions among spatially extended LINER-like galaxies.

%% Please use the acknowledgment and contribution environments. This will 
%% be anonomyized when the "anonymous" style option is used. 
\begin{acknowledgments}
We thank the anonymous referee for a careful and constructive reading of the manuscript, 
whose comments substantially improved the content and completeness of this work.

% individuals
KO acknowledges support from the Korea Astronomy and Space Science Institute under the R\&D program (Project No. 2026-1-831-01), 
supervised by the Korea AeroSpace Administration, 
and the National Research Foundation of Korea (NRF) grant funded by the Korea government (MSIT) (RS-2025-00553982).
GQ and OÇ are supported by an Australian Government Research Training Program (RTP) Scholarship (https://doi.org/10.82133/C42F-K220).
JHL acknowledges support from the National Research Foundation of Korea (NRF) grant funded by the Korea government (MSIT) (No. 2022R1A2C1004025).
AR recognizes the support from the Australian Research Council Centre of Excellence in Optical Microcombs for Breakthrough Science (project number CE230100006), funded by the Australian Government.
SMS acknowledges funding from the Australian Research Council (DE220100003). Parts of this research were conducted by the Australian Research Council Centre of Excellence for All Sky Astrophysics in 3 Dimensions (ASTRO 3D), through project number CE170100013.
SO acknowledges support from the Korean NRF (RS-2023-00214057;RS-2025-00514475).
ST acknowledges the support from the Royal Thai Government Scholarship and the University of Sydney Postgraduate Research Supplementary Scholarship.

% Hector
The Hector Galaxy Survey is based on observations made at the Anglo-Australian Telescope. 
We acknowledge the traditional owners of the land on which the AAT stands, the Gamilaraay people, 
and pay our respects to elders past and present. The Hector multi-object integral field spectrograph instrument was built 
jointly by the University of Sydney and Macquarie University nodes of the Astralis Astronomical Instrumentation Consortium (https://astralis.org.au/), 
with additional financial contributions from the Australian National University and 
University of Western Australia and supported by the Australian Research Council through grants LE170100242, LE190100018 and FT180100231. 
The Hector input catalogue is based on data taken from the WAVES Survey, Sloan Digital Sky Survey, GAMA Survey, 2dfGRS and Skymapper Southern Sky Survey. 
The Hector Galaxy Survey research was supported by the Australian Research Council Centre of Excellence for All Sky Astrophysics in 3 Dimensions (ASTRO3D), 
through project number CE170100013, and other participating institutions. 
The Hector Galaxy Survey website is \url{https://hector.survey.org.au/}. 
The Hector Galaxy Survey makes use of Data Central services (datacentral.org.au).

% Chandra
The scientific results reported in this article are based in part on data obtained from the Chandra Data Archive.
This paper employs a list of Chandra datasets, obtained by the Chandra X-ray Observatory, contained in the Chandra
Data Collection (CDC) at \dataset[doi:10.25574/cdc.521]{https://doi.org/10.25574/cdc.521}

% NED
This research has made use of the NASA/IPAC Extragalactic Database (NED),
which is operated by the Jet Propulsion Laboratory, California Institute of Technology,
under contract with the National Aeronautics and Space Administration.

\end{acknowledgments}

\software{XSPEC \citep{Arnaud96}, 
\texttt{MGEfit} \citep{Cappellari02}, 
pPXF \citep{Cappellari04}, 
CIAO \citep{Fruscione06}, 
\texttt{kcorrect} \citep{Blanton07}, 
\texttt{LZIFU} \citep{Ho16}, 
\texttt{Spaxelsleuth} \citep[][P. K. Das in prep.]{Zovaro24}, 
\texttt{photutils} \citep{Bradley25}}


\appendix

\section{Optical spectral fits} \label{sec:bestfit}
\begin{figure*}
\centering
\includegraphics[height=0.84\textheight, angle=0]{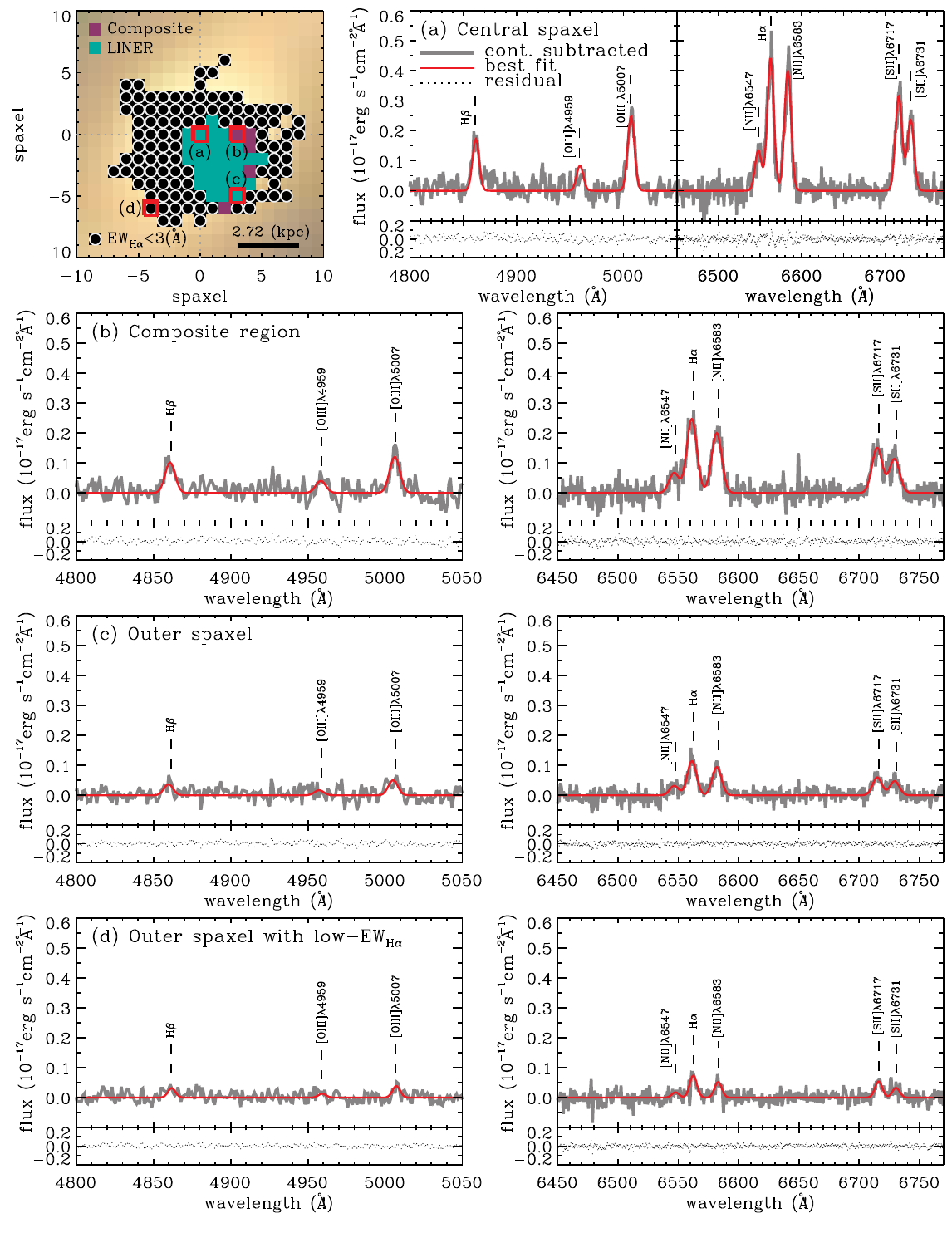}
\caption{Example spectra extracted from representative spaxels, shown with best-fit models. 
The top-left panel displays the spatially resolved classification map 
from the \NIIHa\ BPT diagnostics, 
overlaid on the \textit{grz} color-composite image cutout, 
as shown in Figure~\ref{fig:hector_bpt}. 
Four selected spaxels marked with red boxes corresponding to:  
(a) the central spaxel, 
(b) a composite region, 
(c) an outer spaxel, and 
(d) a region with low-${\rm EW}_{\rm H\alpha}$.  
For each spaxel, the \Hb\ and \Ha\ spectral regions are shown. 
In each panel, the continuum-subtracted spectrum is plotted in grey, 
with the best-fit model overlaid in red. Residuals are shown beneath each spectrum.  
The centroids of key emission lines (\Hb, \OIIIab, \NIIab, \Ha, and \SII) are indicated by vertical dashed lines. 
All spectral panels share the same vertical (flux) scale to facilitate direct comparison. 
\label{fig:bestfit}}
\end{figure*}

Figure~\ref{fig:bestfit} presents example spectra extracted from representative spaxels in the Hector data cube, 
overlaid with the corresponding best-fit emission-line models. Four regions were selected to illustrate typical spectral properties across the field: 
the central spaxel, a composite region, an outer spaxel, and an outer region with low ${\rm EW}_{\rm H\alpha}$. 
These locations are marked by red boxes in the emission diagnostic map.
For each spaxel, we show the continuum-subtracted spectrum along with the fitted emission-line model. 
The displayed examples highlight the quality of the fits and the variation in line strengths across different regions of the galaxy.

% ------------------------------------------------------------------------------------------------------------------------------------------------------------------
 % -------------------------------------------------------------------------------------------------------------------------------------------------

\section{Radiative shock model comparison}
\label{sec:shock_models}

As a qualitative complement to the empirical line-ratio--$\sigma_{\rm gas}$ tests in Section~\ref{subsec:shock},
we compare the spatially resolved \NIIHa\ BPT measurements to the radiative shock model grids of \citet{Allen08}.
These models were computed with the MAPPINGS~III shock and photoionization code and predict optical emission-line
ratios for fast radiative shocks, including both (i) ``shock-only'' models and (ii) ``shock+precursor'' models in
which a photoionized precursor contributes additional high-ionization emission.
The grids span shock velocity and magnetic parameter ($B/\sqrt{n}$ in the \citealt{Allen08} formalism)
for different pre-shock gas densities.
For consistency with the observational definition used for the Hector spaxels, we compute the model grids in terms of
$\log([\mathrm{N\,II}]\,\lambda6583/\mathrm{H\alpha})$.

\begin{figure*}
\centering
\includegraphics[width=1.0\linewidth]{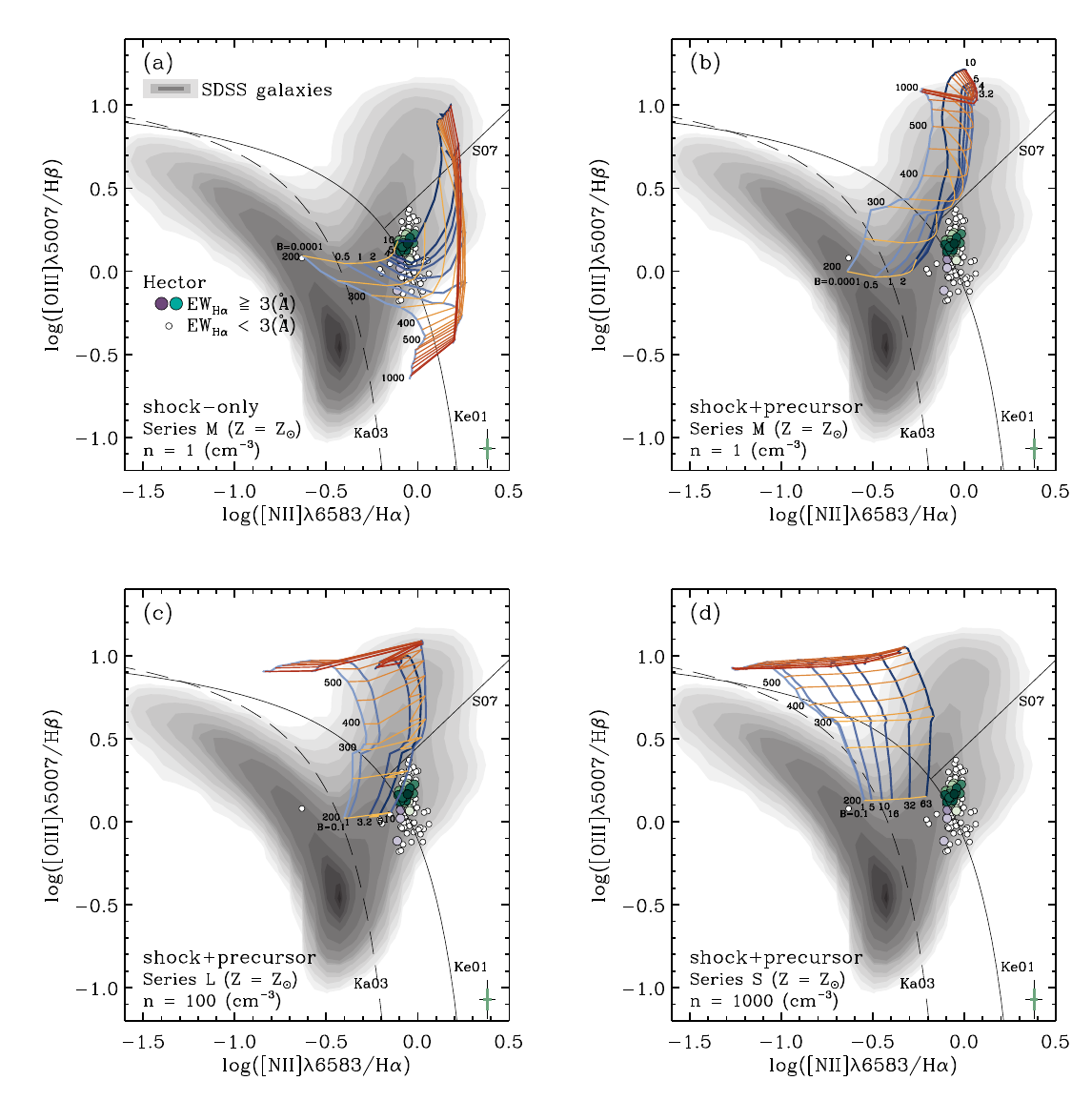}
\caption{The \NIIHa\ BPT diagnostic diagram with radiative shock model grids from \citet{Allen08}.
SDSS background contours show OSSY galaxies \citep{Oh11}.
Hector spaxels with ${\rm S/N} \geq 3$ in \NII, H$\alpha$, \OIII, and H$\beta$ are overplotted;
spaxels with ${\rm EW}_{\rm H\alpha}<3$\,\AA\ are shown as small open symbols, while
spaxels with ${\rm EW}_{\rm H\alpha}\ge3$\,\AA\ are highlighted with larger symbols and their \NIIHa-based classes
(Composite and LINER). See Figure~\ref{fig:hector_bpt} for the same spaxel selection and class definition.
Panels show: (a) $n=1~{\rm cm^{-3}}$ shock-only models, (b) $n=1~{\rm cm^{-3}}$ shock+precursor models,
(c) $n=100~{\rm cm^{-3}}$ shock+precursor models, and (d) $n=1000~{\rm cm^{-3}}$ shock+precursor models.
Blue curves trace constant magnetic parameter $B/n^{1/2}$ and orange curves trace constant shock velocity, as in \citet{Allen08}.
\label{fig:allen08_grid}}
\end{figure*}

Figure~\ref{fig:allen08_grid} shows the \NIIHa\ BPT diagnostic diagram with four representative model configurations.
Panel~(a) presents the solar-abundance shock-only grid for $n=1~{\rm cm^{-3}}$,
while panel~(b) shows the corresponding $n=1~{\rm cm^{-3}}$ shock+precursor grid.
Panels~(c) and (d) show shock+precursor grids for higher pre-shock densities,
$n=100~{\rm cm^{-3}}$ and $n=1000~{\rm cm^{-3}}$, respectively.
These panels correspond to the \citet{Allen08} model set adopted in this work
(series labels M/L/S in our implementation; see \citealt{Allen08} for the detailed abundance prescriptions).

Hector spaxels are required to satisfy ${\rm S/N} \geq 3$ in all four diagnostic lines
(\NII, H$\alpha$, \OIII, and H$\beta$).
For visualization, spaxels with ${\rm EW}_{\rm H\alpha}<3$\,\AA\ are shown as small white filled dots,
while spaxels with ${\rm EW}_{\rm H\alpha}\ge3$\,\AA\ are highlighted with larger colored symbols and their \NIIHa-based classes,
consistent with Figure~\ref{fig:hector_bpt}.
The line ratios shown here are based on single-component Gaussian fits; if multiple kinematic
components are present, such one-component measurements can dilute component-specific excitation signatures
(e.g., a shock-driven contribution confined to a broader, lower-flux component).

In panel~(a) (shock-only, series~M with $n=1~\mathrm{cm^{-3}}$), the higher-EW spaxels
(${\rm EW}_{\rm H\alpha}\ge3$\,\AA) show qualitative overlap with the model grid at moderate shock velocities
($v_{\rm shock}\sim 200$--$250~\mathrm{km\,s^{-1}}$) and magnetic parameters of order
$B/\sqrt{n}\sim$ a few to $\sim10$ in the \citealt{Allen08} notation.
The lower-EW spaxels (${\rm EW}_{\rm H\alpha}<3$\,\AA) occupy a broader range in $\log([\mathrm{O\,III}]/\mathrm{H\beta})$
at comparable $\log([\mathrm{N\,II}]/\mathrm{H\alpha})$, spanning both above and below the locus traced by the higher-EW
population while remaining largely within the overall \NIIHa\ range covered by this model set.
In panel~(b) (shock+precursor, series~M with
$n=1~\mathrm{cm^{-3}}$), the precursor contribution
shifts the grid toward higher
$\log([\mathrm{O\,III}]/\mathrm{H\beta})$ relative to
the shock-only case in panel~(a), placing most of the
grid above the observed spaxel locus.
In panel~(c) (shock+precursor, series~L with $n=100~\mathrm{cm^{-3}}$), the model locus is generally shifted toward
higher $\log([\mathrm{O\,III}]/\mathrm{H\beta})$ than the observed spaxels. The Hector points approach the grid only
near its lower envelope, where the effective contribution of the precursor is smallest within this model grid.
In panel~(d) (shock+precursor, series~S with
$n=1000~\mathrm{cm^{-3}}$), the grid shifts toward
lower $\log([\mathrm{O\,III}]/\mathrm{H\beta})$
compared to panel~(c), but still does not overlap the
bulk of the observed spaxels.
Taken at face value, panels~(a) and (c) therefore indicate that radiative-shock models can reproduce the observed
\NIIHa\ ratios in a phenomenological sense, but only over a restricted part of the available parameter space.

We emphasize, however, that proximity in BPT space does not uniquely imply shock-dominated excitation.
The \citet{Allen08} grids are intrinsically degenerate in abundance/metallicity, density, magnetic parameter,
shock velocity, and precursor contribution, so distinct combinations can map to similar \NIIHa\ ratios.
Moreover, LINER-like line ratios can also arise from photoionization by a hard continuum at low ionization parameter,
which is independently supported for our target by the X-ray analysis in the main text.
Accordingly, we use Figure~\ref{fig:allen08_grid} as an orientation tool showing where the spaxels fall relative to
canonical radiative-shock loci in the \NIIHa\ BPT plane, rather than as a definitive shock diagnostic.
Our physical assessment of shock dominance relies primarily on the absence of the expected low-ionization
line-ratio--$\sigma_{\rm gas}$ behavior after controlling for projected radius
(Figure~\ref{fig:shock_check}), together with the independent multi-wavelength evidence for an LLAGN and the
ionization-parameter trends inferred from our spatially resolved diagnostics.

% -------------------------------------------------------------------------------------------------------------------------------------------------
 % -------------------------------------------------------------------------------------------------------------------------------------------------

\section{\texorpdfstring{[S\,{\sc ii}]}{[S II]} doublet ratio as a gas-pressure diagnostic}
\label{sec:sii_ratio}

As an additional check on potential shock contributions, we examined the spatial distribution of the
[S\,II] $\lambda6717/\lambda6731$ doublet ratio, which is sensitive to the electron density in the warm ionized gas
and therefore provides an empirical proxy for gas pressure at fixed temperature.
We define $R \equiv F(\lambda6717)/F(\lambda6731)$ and measure $R$ on a spaxel-by-spaxel basis using single-component
Gaussian fluxes from the Hector data cube.
Spaxels are required to satisfy ${\rm S/N}(\lambda6717)\ge 3$, ${\rm S/N}(\lambda6731)\ge 3$, and ${\rm S/N}({\rm H}\alpha)\ge 3$.
Within these cuts, 210 spaxels have valid $R$ measurements.

The $R$ distribution spans $R_{\rm min}=0.748$ to $R_{\rm max}=2.620$, 
with a median of $R_{\rm med}=1.358$ 
and percentiles $R_{16}=1.182$ and $R_{84}=1.574$. 
The theoretical low-density limit at $T_{e} = 10^{4}$\,K is
$R \simeq 1.44$ \citep{Osterbrock06}. 
Individual values above this limit are not physically realizable in a single-temperature, 
single-phase medium and instead reflect the propagated uncertainty on the doublet ratio. 
At the adopted detection threshold of S/N $\geq 3$ on each doublet line, 
simple Gaussian error propagation gives $\sigma_{R}/R \gtrsim \sqrt{2}/3 \simeq0.47$, 
so excursions to $R \simeq 2$ at fixed S/N are consistent with random measurement scatter 
rather than physical density variation.

\begin{figure*}
\centering
\includegraphics[width=0.9\linewidth]{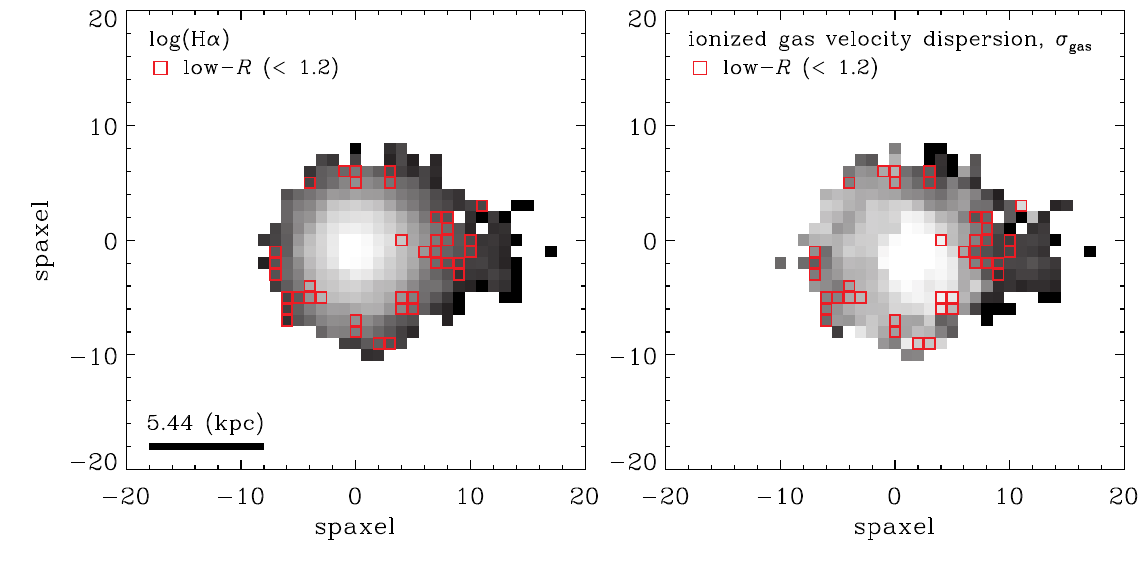}
\caption{Spatial comparison of the [S\,II] doublet-ratio selection with the emission-line morphology and gas kinematics.
Left: $\log F({\rm H}\alpha)$ map (arbitrary flux scaling; shown only for relative morphology), 
displayed with a linear grayscale stretch between the 5th and 99th percentiles of the valid spaxels. 
Right: ionized-gas velocity-dispersion map, $\sigma_{\rm gas}$, 
shown with a linear grayscale stretch between the 5th and 95th percentiles.
In both panels, lighter shades indicate higher values and darker shades indicate lower values.
Red squares mark spaxels with low [S\,II] doublet ratio, $R \equiv F(\lambda6717)/F(\lambda6731) < 1.2$, 
measured from single-component Gaussian fluxes with ${\rm S/N}(\lambda6717)\ge 3$, 
${\rm S/N}(\lambda6731)\ge 3$, and ${\rm S/N}({\rm H}\alpha)\ge 3$.
The low-$R$ spaxels appear in localized patches rather than forming a coherent high-dispersion structure, 
and they are not associated with enhanced $\sigma_{\rm gas}$ in our data, 
consistent with radiative shocks not being the dominant ionization source of the extended LINER-like emission.
A physical scale bar is shown in the left panel.}
\label{fig:sii_ratio_overlay}
\end{figure*}

To isolate spaxels whose doublet ratio is suggestive of genuine density enhancement, 
we adopt the threshold $R < 1.20$. 
Using PyNeb \citep{Luridiana15} with the [S\,{\sc ii}] atomic data adopted therein, 
this threshold corresponds to an electron density of $n_{e} \simeq 3 \times 10^{2}$\,cm$^{-3}$ 
at $T_{e} = 10^{4}$\,K (varying by less than a factor of $\sim 1.6$ over $T_{e} = 5000$--$20000$\,K), 
well below the critical densities of the [S\,{\sc ii}] doublet 
($n_{c} \simeq 1.5 \times 10^{3}$ and $3.9 \times 10^{3}$\,cm$^{-3}$\, for 
$\lambda 6717$ and $\lambda 6731$, respectively; \citealt{Vaona12}). 
The cut selects 39 of 210 spaxels (18.6\%) and highlights several localized patches 
across the emission-line footprint (Figure~\ref{fig:sii_ratio_overlay}).

% -------------------------------------------------------------------------------------------------------------------------------------------------
% -------------------------------------------------------------------------------------------------------------------------------------------------

However, we caution that the inferred density/pressure depends on temperature and on the assumed single-phase interpretation,
and $R$ alone is not a unique shock diagnostic.
In our data, the $R<1.2$ spaxels are not associated with enhanced gas kinematics:
their velocity dispersion is in fact lower than that of the complementary set,
with $\sigma_{\rm gas}$ median $108~{\rm km~s^{-1}}$ (16--84\%: 87--147~${\rm km~s^{-1}}$) for $R<1.2$
versus $141~{\rm km~s^{-1}}$ (16--84\%: 113--172~${\rm km~s^{-1}}$) for $R\ge 1.2$.
Likewise, the $R<1.2$ spaxels do not show enhanced H$\alpha$ surface brightness:
their median $\log F({\rm H}\alpha)$ is lower than that of the $R\ge 1.2$ subset.
These behaviors are not what would be expected if the low-$R$ regions traced widespread radiative shocks that dominate the
LINER-like excitation, where one typically anticipates elevated line widths and/or enhanced emission coincident with the shocked gas.

To verify that this contrast is not driven by spaxels with above-theoretical-limit ratios, 
we further partition the high-$R$ sample into two sub-bins: 
a physically valid sub-bin ($1.20 \leq R \leq 1.44$, $N = 103$) and an above-limit sub-bin ($R > 1.44$, $N = 68$). 
The two sub-bins yield indistinguishable $\sigma_{\rm gas}$ distributions, 
both with median $\sigma_{\rm gas} \simeq 141$\,km\,s$^{-1}$ 
(two-sided Mann--Whitney $p = 0.13$, Kolmogorov--Smirnov $p = 0.26$). 
The elevated median of the $R \geq 1.20$ sample is therefore 
not produced by noise-driven excursions above the theoretical limit. 
The contrast between $R < 1.20$ and $1.20 \leq R \leq 1.44$ remains highly significant, 
with a median offset of $\Delta \sigma_{\rm gas, med} = 32$\,km\,s$^{-1}$ 
(Mann--Whitney $p = 1.7 \times 10^{-4}$, K--S $p = 5.4 \times 10^{-4}$). 
The original two-bin contrast is therefore robust against the inclusion of above-limit spaxels. 
The low-$R$ spaxels are spatially extended, spanning a bounding box of $\sim 10 \times 9$\,kpc, 
about 3.5 times the PSF FWHM, which rules out a kinematically cold nuclear component within an unresolved PSF 
as the origin of their lower $\sigma_{\rm gas}$.

We caution that the [S\,{\sc ii}]\,$\lambda\lambda$6716,6731 doublet is collisionally saturated 
at $n_{e} \gtrsim 10^{3}$--$10^{4}$\,cm$^{-3}$ and is therefore 
intrinsically insensitive to compressed, 
high-density gas phases that may be present in the cooling zones behind fast radiative shocks 
or in clumpy AGN-driven outflows \citep{Kakkad18, Baron19}. 
The kinematic comparison presented here therefore constrains shock-driven turbulence only in the diffuse, 
low-density medium to which the [S\,{\sc ii}] doublet is sensitive. 
A localized contribution of fast shocks confined to gas above the [S\,{\sc ii}] critical density 
cannot be excluded by the present analysis and would require higher-critical-density tracers 
(e.g., [Ar\,{\sc iv}]\,$\lambda\lambda$4711,4740; \citealt{Binette24}) to be probed directly.

Overall, the spatially patchy nature of the low-$R$ regions, 
their lack of correspondence with elevated $\sigma_{\rm gas}$, 
and the robustness of this result against measurement-noise contamination together 
support the main conclusion of Section~\ref{subsec:shock}. 
Radiative shocks are unlikely to be the dominant ionization source of the
extended LINER-like emission in the diffuse gas to which the [S\,{\sc ii}] diagnostic is sensitive. 
Localized compression and pressure enhancements at densities above the [S\,{\sc ii}] saturation limit, 
whether driven by internal dynamics, AGN outflows, or the cluster environment, cannot be excluded by the present analysis.

%% For this sample we use BibTeX plus aasjournalv7.bst to generate the

\bibliography{reference}{}
\bibliographystyle{aasjournalv7}



\end{document}